\documentclass[twocolumn]{aastex63}
\usepackage{amsmath}
\usepackage{natbib}
\usepackage{float}
\usepackage{CJKutf8} 
\usepackage{placeins}
\usepackage{mhchem} 

\newcommand{\teff}{$T_\mathrm{eff}$}
\newcommand{\logg}{$\log{g}$}

\newcommand{\vsini}{$v\sin{i}$}
\newcommand{\kms}{km~s$^{-1}$}

\newcommand{\caltech}{Department of Astronomy, California Institute of Technology, Pasadena, CA 91125, USA}
\newcommand{\gps}{Division of Geological \& Planetary Sciences, California Institute of Technology, Pasadena, CA 91125, USA}
\newcommand{\ucsc}{Department of Astronomy \& Astrophysics, University of California, Santa Cruz, CA95064, USA}
\newcommand{\keck}{W. M. Keck Observatory, 65-1120 Mamalahoa Hwy, Kamuela, HI, USA}
\newcommand{\ucla}{Department of Physics \& Astronomy, 430 Portola Plaza, University of California, Los Angeles, CA 90095, USA}
\newcommand{\jpl}{Jet Propulsion Laboratory, California Institute of Technology, 4800 Oak Grove Dr.,Pasadena, CA 91109, USA}
\newcommand{\ucsd}{Department of Astronomy \& Astrophysics,  University of California, San Diego, La Jolla, CA 92093, USA}

\newcommand{\osu}{Department of Astronomy, The Ohio State University, 100 W 18th Ave, Columbus, OH 43210 USA}

\newcommand{\arizona}{James C. Wyant College of Optical Sciences, University of Arizona, Meinel Building 1630 E. University Blvd., Tucson, AZ. 85721}

\submitjournal{ApJL}

\shorttitle{PDS 70b Has Stellar-like C/O}
\shortauthors{Hsu et al.}

\graphicspath{{./}{figures/}}

\begin{document}

\title{PDS 70b Shows Stellar-like Carbon-to-oxygen Ratio}

\correspondingauthor{Chih-Chun Hsu}
\email{chsu@northwestern.edu}

\author[0000-0002-5370-7494]{Chih-Chun Hsu}
\affil{Center for Interdisciplinary Exploration and Research in Astrophysics (CIERA), Northwestern University,
1800 Sherman Ave, Evanston, IL, 60201, USA}

\author[0000-0003-0774-6502]{Jason J. Wang
\begin{CJK*}{UTF8}{gbsn}(王劲飞)\end{CJK*}}
\affil{Center for Interdisciplinary Exploration and Research in Astrophysics (CIERA), Northwestern University,
1800 Sherman Ave, Evanston, IL, 60201, USA}
\affil{Department of Physics and Astronomy, Northwestern University, 2145 Sheridan Rd, Evanston, IL 60208, USA}

\author[0000-0003-0787-1610]{Geoffrey A. Blake}
\affiliation{\gps}

\author[0000-0002-6618-1137]{Jerry W. Xuan}
\affiliation{\caltech}

\author[0000-0003-0097-4414]{Yapeng Zhang}
\affiliation{\caltech}

\author[0000-0003-2233-4821]{Jean-Baptiste Ruffio}
\affiliation{\ucsd}

\author[0000-0001-9708-8667]{Katelyn Horstman}
\affiliation{\caltech}

\author[0000-0003-1172-5755]{Julianne Cronin}
\affil{Center for Interdisciplinary Exploration and Research in Astrophysics (CIERA), Northwestern University,
1800 Sherman Ave, Evanston, IL, 60201, USA}
\affil{Department of Physics and Astronomy, Northwestern University, 2145 Sheridan Rd, Evanston, IL 60208, USA}

\author[0000-0003-1399-3593]{Ben Sappey}
\affiliation{\ucsd}

\author[0000-0002-6171-9081]{Yinzi Xin}
\affiliation{\caltech}

\author[0000-0002-1392-0768]{Luke Finnerty}
\affiliation{\ucla}

\author[0000-0002-1583-2040]{Daniel Echeverri}
\affiliation{\caltech}

\author[0000-0002-8895-4735]{Dimitri Mawet}
\affiliation{\caltech}
\affiliation{\jpl}

\author[0000-0001-5213-6207]{Nemanja Jovanovic}
\affiliation{\caltech}

\author[0000-0001-5173-2947]{Clarissa R. Do Ó}
\affiliation{\ucsd}


\author[0000-0002-6525-7013]{Ashley Baker}
\affiliation{\caltech}

\author{Randall Bartos} 
\affiliation{\jpl}

\author[0000-0003-4737-5486]{Benjamin Calvin}
\affiliation{\caltech}
\affiliation{\ucla}

\author{Sylvain Cetre} 
\affiliation{\keck}

\author[0000-0001-8953-1008]{Jacques-Robert Delorme}
\affiliation{\keck}

\author{Gregory W. Doppmann} 
\affiliation{\keck}

\author[0000-0002-0176-8973]{Michael P. Fitzgerald}
\affiliation{\ucla}

\author[0000-0002-4934-3042]{Joshua Liberman}
\affiliation{\arizona}

\author[0000-0002-2019-4995]{Ronald A. L\'opez}
\affiliation{\ucla}

\author[0000-0003-3165-0922]{Evan Morris}
\affiliation{\ucsc}

\author{Jacklyn Pezzato-Rovner} 
\affiliation{\caltech}

\author{Tobias Schofield}
\affiliation{\caltech}

\author[0000-0001-6098-3924]{Andrew Skemer}
\affiliation{\ucsc}

\author[0000-0001-5299-6899]{J. Kent Wallace}
\affiliation{\jpl}

\author[0000-0002-4361-8885]{Ji Wang \begin{CJK*}{UTF8}{gbsn}(王吉)\end{CJK*}}
\affiliation{\osu}


\begin{abstract}
The $\sim$5~Myr PDS 70 is the only known system with protoplanets residing in the cavity of the circumstellar disk from which they formed, ideal for studying exoplanet formation and evolution within its natal environment.
Here we report the first spin constraint and C/O measurement of PDS 70b from Keck/KPIC high-resolution spectroscopy. 
We detected CO (3.8~$\sigma$) and H$_2$O (3.5~$\sigma$) molecules in the PDS 70b atmosphere via cross-correlation, with a combined CO and H$_2$O template detection significance of 4.2~$\sigma$.
Our forward model fits, using BT-Settl model grids, provide an upper limit for the spin-rate of PDS 70b ($<$29~{\kms}). 
The atmospheric retrievals constrain the PDS 70b C/O ratio to ${0.28}^{+0.20}_{-0.12}$ ($<${0.63} under 95\% confidence level) and a metallicity [C/H] of ${-0.2}^{+0.8}_{-0.5}$~dex, consistent with that of its host star.
The following scenarios can explain our measured C/O of PDS 70b in contrast with that of the gas-rich outer disk (for which C/O $\gtrsim$ 1).
First, the bulk composition of PDS 70b might be dominated by dust+ice aggregates rather than disk gas.
Another possible explanation is that the disk became carbon-enriched {\it after} PDS 70b was formed, as predicted in models of disk chemical evolution and as observed in both very low mass star and older disk systems with \textit{JWST}/MIRI.
Because PDS 70b continues to accrete and its chemical evolution is not yet complete, more sophisticated modeling of the planet and the disk, and higher quality observations of PDS 70b (and possibly PDS 70c), are necessary to validate these scenarios.

\end{abstract}


\section{Introduction} \label{sec:intro}

Since the discovery of the first exoplanet around a Sun-like star \citep{Mayor:1995aa}, $>$5,700 are now known\footnote{NASA Exoplanet Archive compilation on 2024 October 7; \url{https://exoplanetarchive.ipac.caltech.edu/}.}.
Among these advances, directly imaged companions offer a unique laboratory to study exoplanet formation and evolution \citep{Bowler:2016aa, Currie:2023ab}.
These imaged planets are mostly young and warm ($<$500~Myr), thus bright, and well separated from their host stars.
Myriad protoplanetary disks have been characterized extensively in the millimeter band, but PDS 70 remains the only system that has confirmed protoplanets b and c residing in the cavity of their natal, gas-rich disk from which they formed \citep{Keppler:2018aa, Muller:2018aa, Haffert:2019aa}\footnote{AB Aur b is a confirmed protoplanet with its disk \citep{Currie:2022aa}. See also other candidates summarized in \cite{Currie:2023ab}.}, serving as a unique opportunity to study exoplanet formation and evolution in situ.

PDS 70 A is a K7 T Tauri star in the Upper Centaurus Lupus association at 112~pc \citep{Pecaut:2016aa, Muller:2018aa, Gaia-Collaboration:2023aa}, with a slightly subsolar metallicity ({[}Fe\slash H{]} = $-0.11 \pm 0.1$) \citep{Steinmetz:2020aa}, a C/O ratio of $0.44 \pm 0.19$ \citep{Cridland:2023aa}, and an age of 5.4 $\pm$ 1.0~Myr \citep{Riaud:2006aa, Pecaut:2016aa}\footnote{We note that PDS 70 A might be a member in the subgroup $\nu$ Cen of the Upper Centaurus Lupus, with reported ages between $\sim$9--16~Myr \citep{Ratzenbock:2023ab, Ratzenbock:2023aa}, while most disk lifetimes are less than 10~Myr \citep{Pfalzner:2024aa}.}. 
PDS 70b has a mass of 2--4~M$_\mathrm{Jup}$, a semi-major axis of 20.8$^{+0.6}_{-0.7}$ au, and a non-zero orbital eccentricity of 0.17 $\pm$ 0.06; while PDS 70c has a mass of 1--3~M$_\mathrm{Jup}$ with a circular orbit (eccentricity 0.037$^{+0.041}_{-0.025}$) at 34.3$^{+2.2}_{-1.8}$ au \citep{Wang:2020aa, Wang:2021ab}. 
PDS 70b and c show clear signs of H$\alpha$ \citep{Haffert:2019aa}, indicative of ongoing accretion, but no detections of H$\beta$ were reported in \cite{Hashimoto:2020aa}.

The PDS 70 outer disk has a gas-phase C/O ratio $\gtrsim$ 1 as inferred from ALMA spectroscopy \citep{Facchini:2021aa, Cridland:2023aa, Law:2024aa}, and the PDS 70 inner disk was detected in H$_2$O and CO$_2$ with \textit{JWST}/MIRI \citep{Perotti:2023aa}.
Both PDS 70b and c lie within the CO ice line (at 56--85~au \citealp{Cridland:2023aa, Law:2024aa}).
The measurements of the C/O ratio of PDS 70b and c are crucial for constraining the formation and evolution of the planetary system, but no clear molecular detections have been reported prior to this work, despite attempts such as the one using the SINFONI integral field spectrograph ($R$ $\sim$ 5000) at the Very Large Telescope \citep{Cugno:2021aa}.
The PDS 70 A and b properties relevant to this work are summarized in Table~\ref{table:system_properties}.

In this Letter, we report the first abundance measurement of the protoplanet PDS 70b using atmospheric molecular line detections of CO and H$_2$O from Keck/KPIC high-resolution spectroscopy.
In Section~\ref{sec:data} we describe our KPIC observation and data reduction.
Section~\ref{sec:ccf_detect} presents our detection of CO and H$_2$O, while
Section~\ref{sec:model} and Section~\ref{sec:retrieve} show our forward-modeling results using a self-consistent modeling and retrieval framework, respectively.
In Section~\ref{sec:discuss} we discuss possible interpretations of our measured C/O of PDS 70b with the host star and disk measurements in the literature.
We summarize our findings in Section~\ref{sec:sum}.

\begin{deluxetable}{lcc}
\tablecaption{PDS 70 A and b Properties \label{table:system_properties}}
\tablecolumns{3}
\tablehead{
\colhead{Property (unit)} &  \colhead{Value}  & \colhead{Ref.}
}
\startdata
\multicolumn{3}{c}{PDS 70~A}\\
\hline
R.A. (J2000) & 14:08:10.15 & (1) \\
Dec. (J2000) & $-$41:23:52.57 & (1) \\
$\mu_{\alpha}$ (mas yr$^{-1}$) & $-29.70 \pm 0.02$ & (1) \\
$\mu_{\delta}$ (mas yr$^{-1}$) & $-24.04 \pm 0.02$ & (1) \\
Mass (M$_{\odot}$) & 0.88 $\pm$ 0.02 & (2) \\
Age (Myr) & $\sim$5 & (3) \\
SpT & K7IVe & (4) \\
\textit{Gaia} $G$ & 11.606$\pm$0.004 & (1) \\
$J_\mathrm{\, MKO}$ (mag) & 9.55 $\pm$ 0.02 & (5) \\
$H_\mathrm{\, MKO}$ (mag) & 8.82 $\pm$ 0.04 & (5) \\
$K_{\rm S, \, MKO}$ (mag) & 8.54 $\pm$ 0.02 & (5) \\
$\pi$ (mas) & 8.898 $\pm$ 0.019 & (1) \\
distance (pc) & 112.4 $\pm$ 0.2 & (1) \\
RV ({\kms})\tablenotemark{a} & 6.65$^{+0.14}_{-0.22}$ & (9) \\
{\vsini} ({\kms}) & 17.3$^{+0.4}_{-0.3}$ & (9) \\
{[}Fe\slash H{]} (dex) & $-0.11 \pm 0.1$ & (6) \\
C\slash O & $0.44 \pm 0.19$ & (7) \\
\hline
\multicolumn{3}{c}{PDS 70b}\\
\hline
Mass (M$_\mathrm{Jup}$) & 2--4 & (7) \\
Radius (R$_\mathrm{Jup}$) & $2.7^{+0.4}_{-0.3}$ & (8) \\
$a$ (au)\tablenotemark{b} & 20.8$^{+0.6}_{-0.7}$ & (2) \\
$e$\tablenotemark{b} & $0.17 \pm 0.06$ & (2) \\
$i$ (deg)\tablenotemark{b} & 130.5$^{+2.5}_{-2.4}$ & (2) \\
{\teff} (K) & 1204$^{+52}_{-53}$ & (8) \\
$L$' (mag) & 14.64 $\pm$ 0.18 & (8) \\
{\vsini} ({\kms}) & $< 29$\tablenotemark{c} & (9) \\
RV ({\kms})\tablenotemark{a} & $-1.7^{+3.4}_{-5.2}$ & (9) \\
{[}C\slash H{]} (dex) & ${-0.2}^{+0.8}_{-0.5}$ & (9) \\
C\slash O & ${0.28}^{+0.20}_{-0.12}$ & (9) \\
\enddata
\tablenotetext{a}{Barycentric RV on MJD 60453.27578}
\tablenotetext{b}{Dynamically stable solutions in \cite{Wang:2021ab}}
\tablenotetext{c}{Upper limit at 95\% confidence level}
\tablerefs{(1) \cite{Gaia-Collaboration:2023aa}; (2) \cite{Wang:2021ab}; (3) \cite{Muller:2018aa}; (4) \cite{Pecaut:2016aa}; (5) \cite{Cutri:2003aa}; (6) \cite{Steinmetz:2020aa}; (7) \cite{Cridland:2023aa}; (8) \cite{Wang:2020aa}; (9) This work}
\end{deluxetable}

\section{Observations \& Data Reduction} \label{sec:data}

The Keck Planet Imager and Characterizer (KPIC) connects the Keck II Telescope's adaptive optics system to the infrared high-resolution spectrograph NIRSPEC \citep{McLean:1998aa, McLean:2000aa, Martin:2018aa}  using single-mode fibers \citep{Mawet:2016aa, Mawet:2017aa, Delorme:2021aa}, enabling diffraction-limited high-resolution spectroscopy ($R \sim$ 35,000).
KPIC underwent a series of upgrades in April 2024 (\citealp{Echeverri:2024aa, Horstman:2024aa}, N. Jovanovic et al., under review).

We observed PDS 70b on 2024 May 23 (UT) under clear and stable weather conditions, with a natural seeing of 0$\farcs$65.
We used KPIC fibers 2 and 4 (out of the four KPIC fibers). 
The total exposure time on PDS 70b was limited to 140~min (fourteen 600~s exposures), due to the source declination and the Nasmyth platform limit of the Keck II telescope. KPIC observations require an orbital prediction of the planet to offset from the host star, and we used the astrometry solutions in \cite{Wang:2021ab} and \texttt{whereistheplanets} \citep{Wang:2021ac} \footnote{\url{http://whereistheplanet.com/}} and offset to the separation of 148.76~mas and position angle of 128.28~deg relative to the host star PDS 70 A during our observations.
We also observed PDS 70 A, and used on-axis star spectra as an empirical star template to model the star light contribution injected into the fiber pointed at the planet.
The exposure time on PDS 70 A was taken as six 180~s exposures using KPIC fibers 2 and 4, for a total of 18 min, roughly before and after a set of 40--60~min exposures on PDS 70b.
For the spectral trace identifications and telluric calibrator, we used the A0V star HD 118214 \citep{Cowley:1969aa}, with two exposures of 30~s for each fiber. 
For the wavelength calibration, we used early M giant star (C-R4IIIb; \citealp{Keenan:1993aa}) HIP 62944, with two exposures of 30~s for each fiber. HIP 62944 has narrow spectral lines and telluric lines that enable precise wavelength calibration.

Our KPIC data were reduced using the \texttt{KPIC Data Reduction Pipeline}\footnote{\url{https://github.com/kpicteam/kpic_pipeline}}, with the procedures detailed in \cite{Wang:2021aa} and \cite{Hsu:2024ab}, using background subtraction and optimal source extraction \citep{Horne:1986aa}.
The PDS 70 A and PDS 70b spectra were background subtracted using the thermal background frames of the corresponding integration times (180~s and 600~s for PDS 70 A and b, respectively). 

\section{Cross-correlation Function Detection} \label{sec:ccf_detect}

\begin{figure*}
    \centering
    \includegraphics[width=0.32\textwidth]{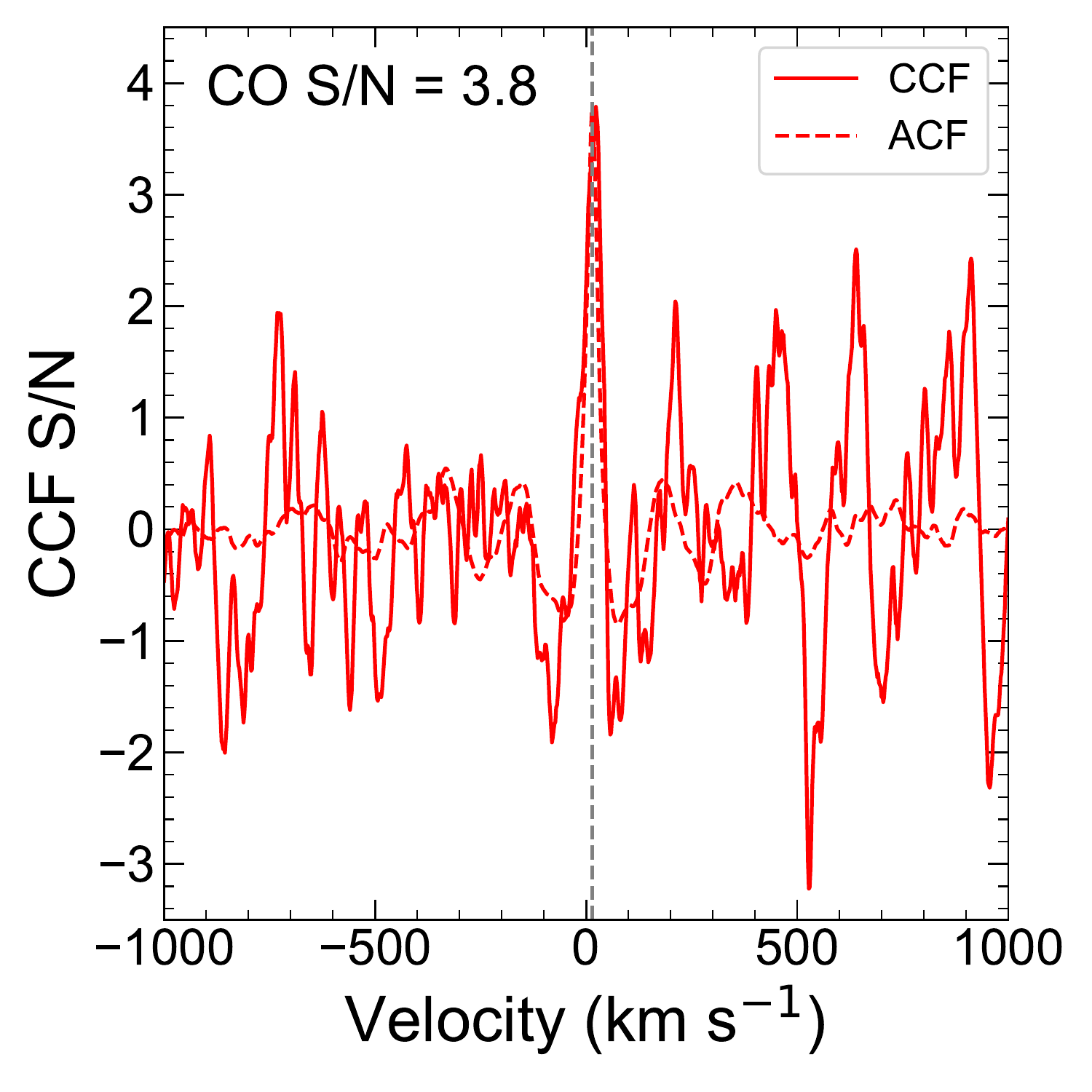}
    \includegraphics[width=0.32\textwidth]{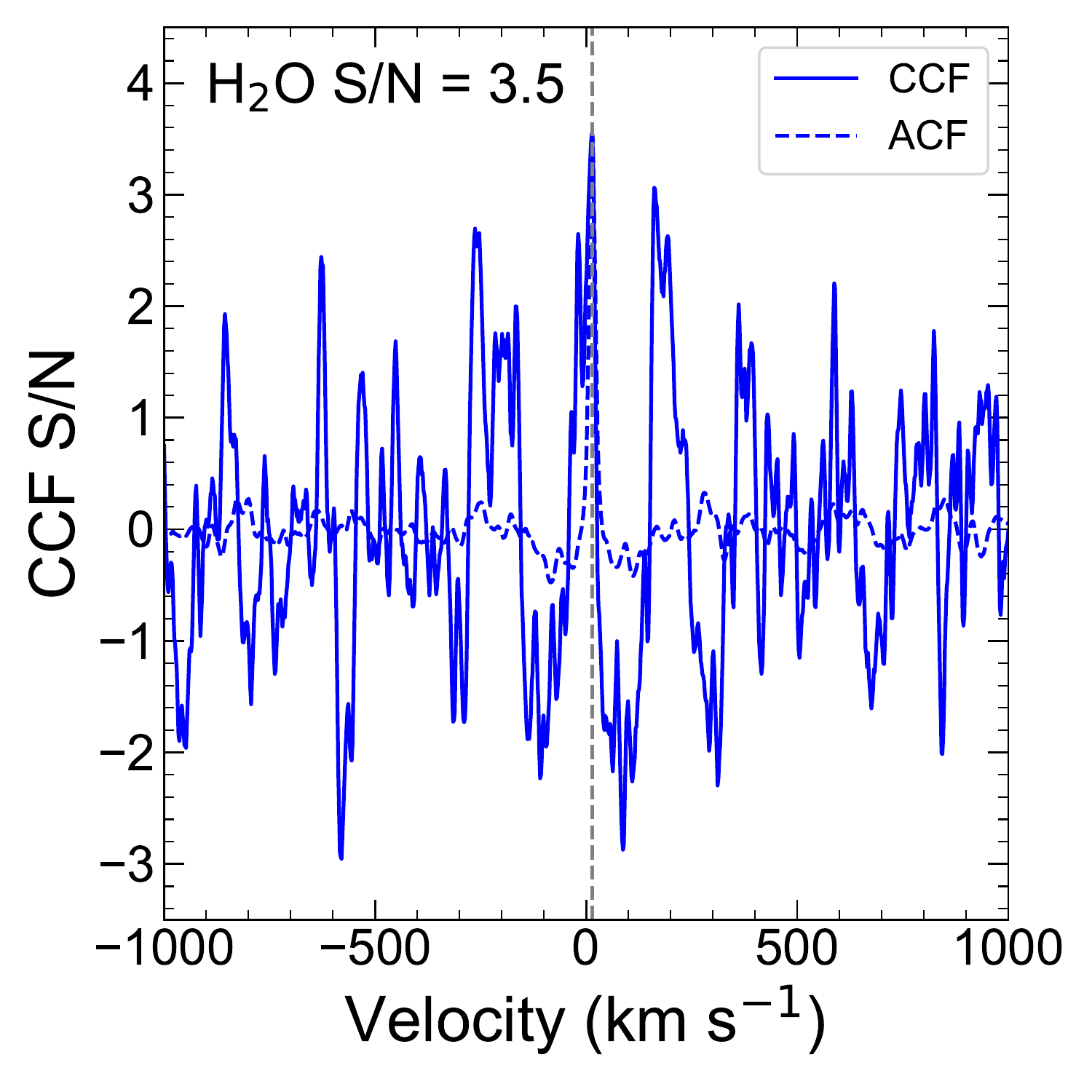}
    \includegraphics[width=0.32\textwidth]{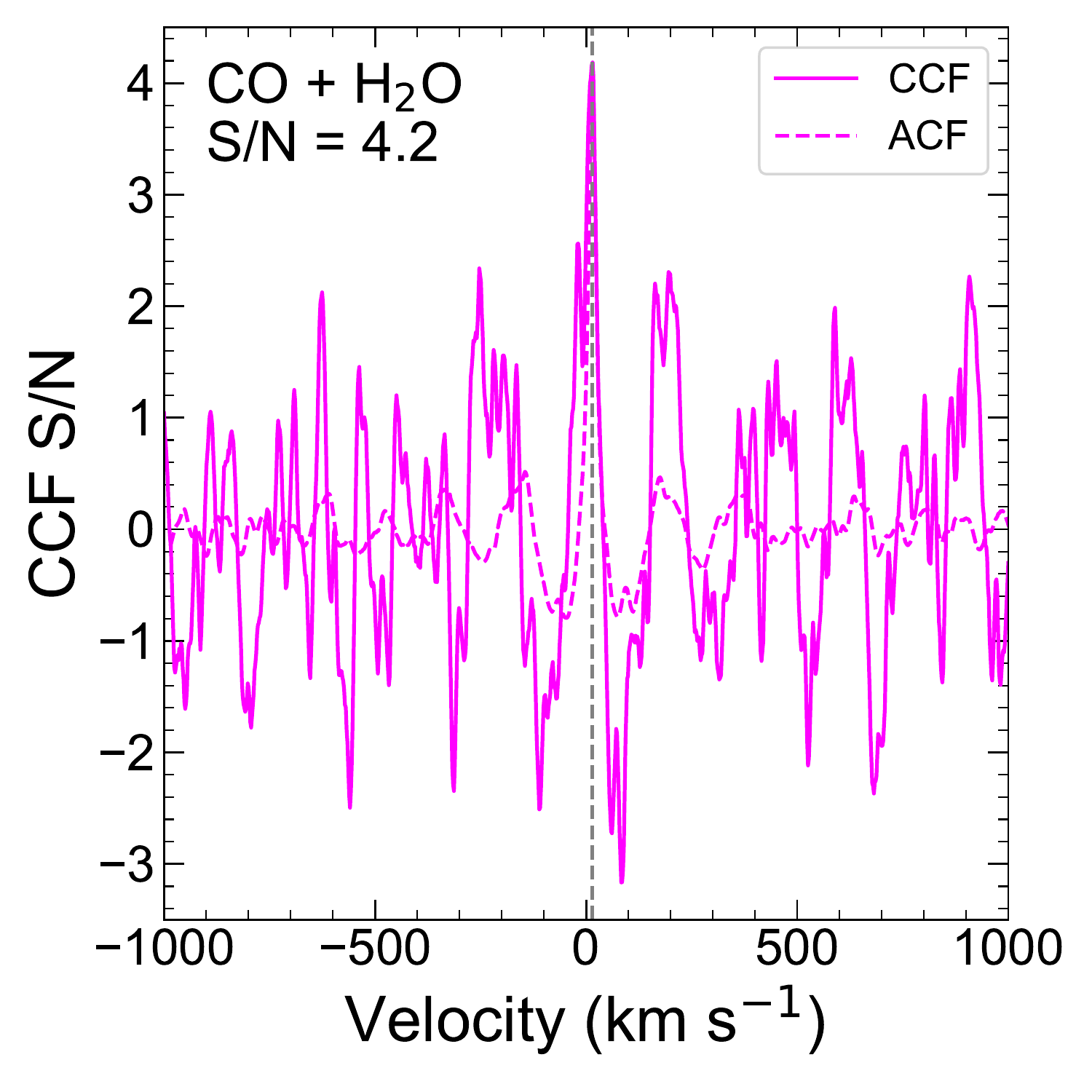}
    \caption{Cross-correlation functions (CCFs) of our PDS 70b KPIC spectra against the molecular templates derived from the Sonora-Bobcat models.
    \textit{Upper}: CCF (red solid line) of our KPIC spectra for the CO molecular templates.
    The stellar barycentric-included RV is depicted by the grey vertical dashed line.
    The auto-correlation function (ACF) of the CO molecular templates, normalized to the peak of CCF, is plotted as a dashed red line. The ACF serves as a guide to detections of given molecular templates.
    \textit{Middle}: Same as the upper panel for the H$_2$O molecular templates in blue.
    \textit{Bottom}: Same as the upper panel for the CO + H$_2$O molecular templates in magenta.
    }
    \label{fig:ccf}
\end{figure*}

Our detection of PDS 70b uses the forward-modeling cross-correlation function (CCF) technique.
We refer readers to \cite{Wang:2021aa, Xuan:2022aa} and \cite{Hsu:2024ab} for the formulation and implementation of our methods.

In short, we jointly fit for the host star contribution, the planet template with various combinations of molecules, and the telluric and instrument response using the least-squares-fit-based CCF.
The CCF signal of a given velocity shift was obtained by optimizing the amplitude of the host star contribution (from the observed on-axis star spectra) and the planet templates.
Our planet templates including only CO or H$_2$O along with combined CO + H$_2$O opacities are derived from Sonora-Bobcat models \citep{Marley:2018aa, Marley:2021aa} at effective {\teff} = 1,200~K and a surface gravity of $\log{g}$ = 4.0~cgs dex \citep{Wang:2021aa}, consistent with the PDS 70b parameters derived in \cite{Wang:2021ab}.
We ran the velocity from $-$1,000 to $+$1,000 {\kms} and used the last $\pm$ 800~{\kms} CCF wings to compute the noise and estimate the signal-to-noise ratio (SNR).
We found detections of CO (CCF SNR $\sim$ 3.8), H$_2$O (CCF SNR $\sim$ 3.5), and CO + H$_2$O (CCF SNR $\sim$ 4.2) in our CCF curves, as shown in Figure~\ref{fig:ccf}.
We further constrain the physical parameters of PDS 70b in Sections~\ref{sec:model} and \ref{sec:retrieve}.

\section{Bulk Parameters of PDS 70 A and b} \label{sec:model}

\subsection{PDS 70 A} \label{sec:host_star_model}

If distinct, fits of the radial velocity (RV) of PDS 70 A and 70 b on the same night can help us validate our protoplanet detection. 
We forward modeled NIRSPEC order 33 (2.29--2.34~$\micron$) of our PDS 70 A spectra using the \texttt{SMART} package \citep{Hsu:2021aa, Hsu:2021ab}, with the procedures detailed in \cite{Hsu:2021aa, Hsu:2023aa}.

The forward method fits the stellar spectra using the BT-Settl CIFIST models \citep{Baraffe:2015aa}, along with a telluric model as a function of airmass and precipitable water vapor \citep{Moehler:2014aa}.
The best-fit parameters were derived using the Markov Chain Monte Carlo (MCMC) package \texttt{emcee} \citep{Foreman-Mackey:2013aa}.
We incorporated a fringe model determined outside of MCMC, using a formulation similar to \cite{Cale:2019aa} (see also \citealp{Xuan:2024aa, Horstman:2024aa}).
There are nine parameters in our MCMC forward model fit: effective temperature ({\teff}), surface gravity ({\logg}), RV, projected rotational velocity ({\vsini}), airmass, precipitable water vapor, and three nuisance parameters (flux and wavelength offsets, and a noise inflation scale factor).
We used 100 chains and 10000 steps, with a burn-in of the first 2000 steps. Convergence occurred within the first 1000 steps verified by visual inspection of the chain evolutions.

Our best-fit values are {\teff} = 3849$^{+108}_{-39}$~K, {\logg} = 5.20 $\pm$ 0.04 cm s$^{-2}$ dex, {\vsini} = 17.3$^{+0.4}_{-0.3}$~{\kms}, and (barycentric velocity corrected) RV = 6.65$^{+0.14}_{-0.22}$~{\kms}, while
our measured {\vsini} is consistent with the literature measurement of 17.16 $\pm$ 0.16~{\kms} \citep{Thanathibodee:2020aa, Swastik:2021aa, Bowler:2023aa}.
Our RV is consistent with lower resolution spectroscopic measurements such as RAVE (4 $\pm$ 7~{\kms}; \citealp{Steinmetz:2020ab}), \textit{Gaia} (3.1 $\pm$ 1.4~{\kms}; \citealp{Gaia-Collaboration:2018ab}), and HARPS RV = $+6.0 \pm 1.5$~{\kms} \citep{Thanathibodee:2020aa}.
Our derived RV of PDS 70 A is also consistent with the optimal RV of 4.8~{\kms} for belonging to the $\nu$ Cen membership reported in \cite{Ratzenbock:2023ab, Ratzenbock:2023aa}, while their RV scatters in the subgroup are typically a few {\kms}.
When determined solely through high-resolution spectroscopy over narrow wavelength ranges, {\teff} and {\logg} measurements can have significant uncertainties. We therefore stress that our RV and {\vsini} constraints, which are our targeted physical parameters in this work, are largely insensitive to the true values of {\teff} and {\logg} (see discussions and justifications in \citealp{Hsu:2021aa, Theissen:2022aa, Hsu:2024aa}).

\subsection{PDS 70b} \label{sec:planet_model}

To determine the atmospheric parameters of PDS 70b, we forward modeled the observed KPIC spectra of PDS 70b by simultaneously fitting stellar, which is the diffracted starlight that leaks into the fiber pointed at the planet, and planet flux contributions. 
The fitting procedures are detailed in \cite{Hsu:2024ab}. 
For the wavelength range of our analysis, we used the NIRSPEC orders 31--33 (2.29--2.49 $\micron$), as these contain CO, H$_2$O, CH$_4$ features and provided the best wavelength calibration.
We used the on-axis KPIC spectra of PDS 70 A to serve as an empirical template and the BT-Settl CIFIST model grids \citep{Baraffe:2015aa} for the planet model. 
In short, we fitted 17 parameters in total, including {\teff}, {\logg}, RV, {\vsini}, planet flux in counts (5 parameters), and stellar contribution along with nuisance parameters (12 parameters). Each order has two star-flux terms in counts (for fibers 2 and 4), one line spread function scale factor, and one noise jitter term.
The best-fit parameters are derived using the nested sampling method \texttt{dynesty} \citep{Speagle:2020aa} with 1000 live points. The normal $\chi^2$-based log-likelihood is defined in \cite{Hsu:2024ab}.
Our priors and resulting parameters are summarized in Table~\ref{table:forward_model_parameters}

\begin{figure*}
    \centering
    \includegraphics[width=\textwidth]{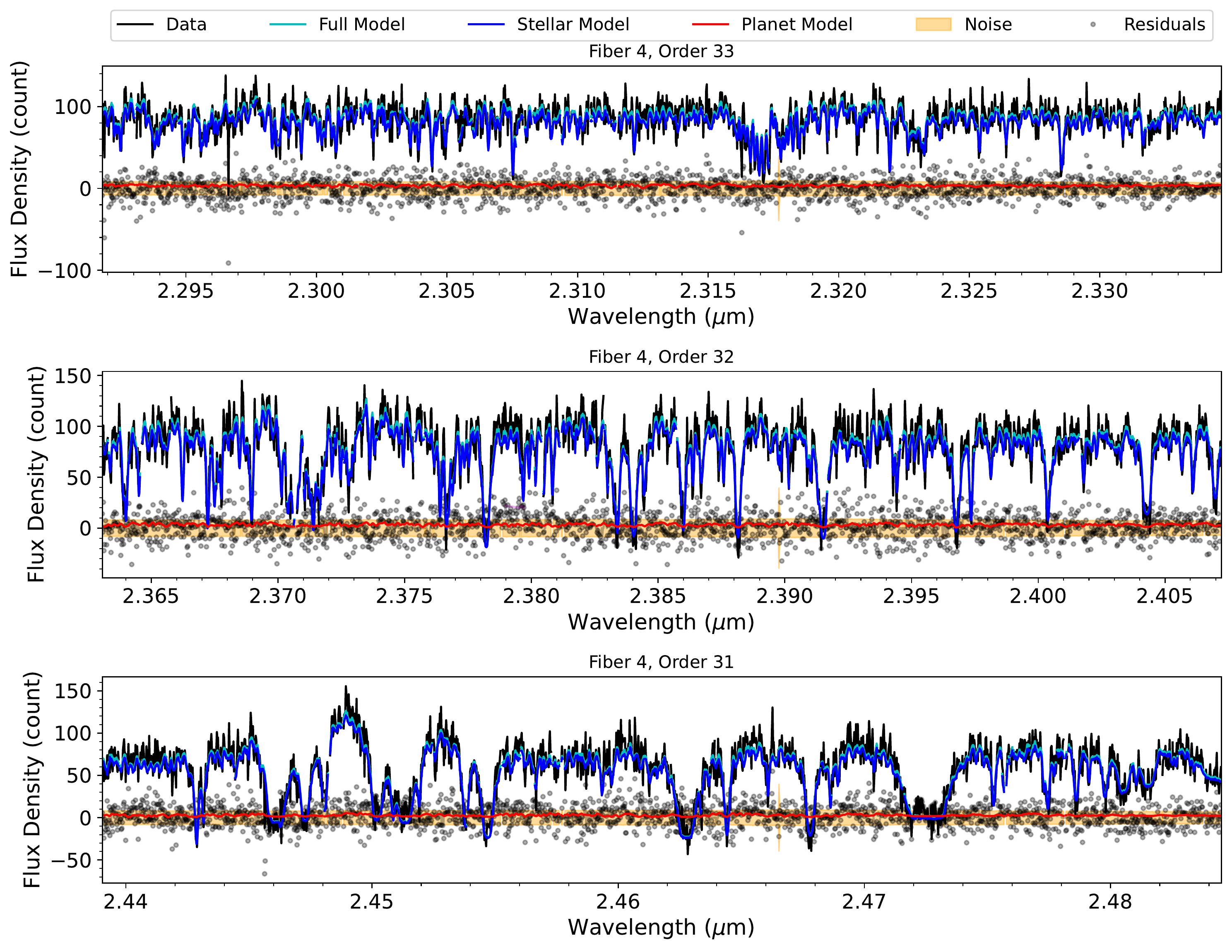}
    \caption{KPIC spectrum of PDS 70b and its best-fit forward model using the BT-Settl model grids.
    The spectra were taken on 2024 May 23 (UT).
    Orders 31--33 on fiber 4 are shown in black lines.
    The full forward model, with the stellar speckle and planet fluxes, is in cyan lines.
    The stellar model, directly from the observed on-axis KPIC spectrum of PDS 70 A, is shown in blue lines. 
    The planet model, including the best-fit BT-Settl model with the observed telluric profile, is illustrated in red lines. 
    The residual (data $-$ full forward model) is depicted in grey dots, and data noise is shown in the orange-shaded region.} 
    \label{fig:pds70b_spectrum_btsettl}
\end{figure*}

\begin{figure}
    \centering
    \includegraphics[width=0.45\textwidth]{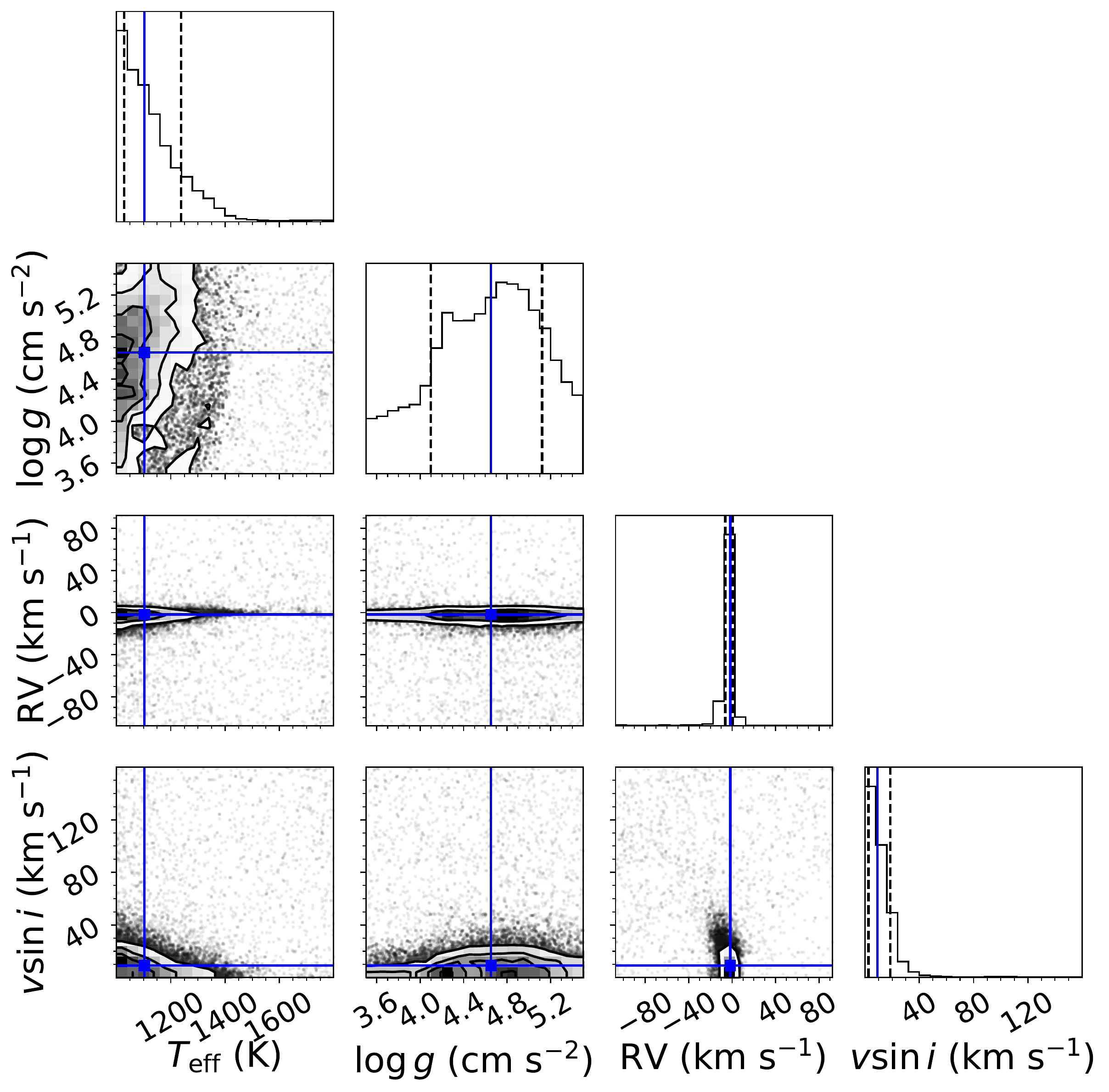}
    \includegraphics[width=0.45\textwidth]{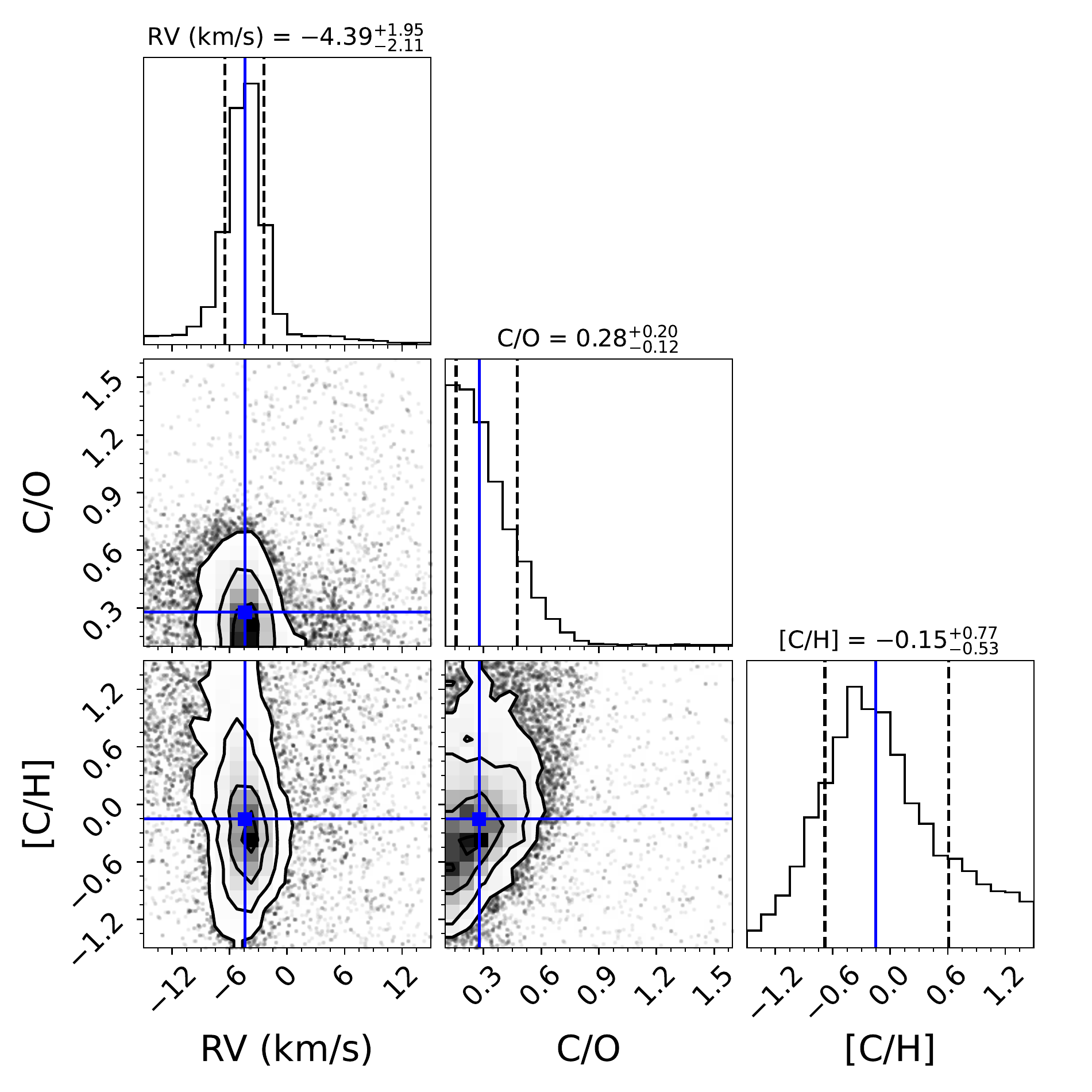}
    \caption{
    \textit{Top}: Posteriors of the derived physical parameters of PDS 70b using the BT-Settl models, including effective temperature, surface gravity, projected rotational and radial velocities. See Section~\ref{sec:planet_model} for details.
    \textit{Bottom}: The posteriors of RV, C/O ratio and metallicity [C/H] derived from KPIC spectra of PDS 70b using the \texttt{petitRADTRANS} retrieval framework. See Section~\ref{sec:retrieve} for details.
    }\label{fig:btsettl_posterior}
\end{figure}

Figure~\ref{fig:pds70b_spectrum_btsettl} shows the KPIC spectra of PDS 70b as well as its best-fit forward model. The posterior probability distributions are shown in Figure~\ref{fig:btsettl_posterior}.
Our best-fit values are {\teff} = 1003$^{+134}_{-75}$~K, {\logg} = 4.7$^{+0.5}_{-0.6}$~cm s$^{-2}$ dex, {\vsini} = 9$^{+9}_{-7}$~{\kms} (consistent with a non-detection of spin, see discussions below), and the barycentric corrected RV = $-1.7^{+2.3}_{-4.6}$~{\kms}.
Our measured PDS 70 A RV is $6.65^{+0.14}_{-0.22}$~{\kms} (Section~\ref{sec:host_star_model}) 
which differs by 3.6~$\sigma$ from PDS 70b (i.e. $\Delta$ RV = RV$_\mathrm{planet}$ - RV$_\mathrm{star}$ = $-8.4^{+2.3}_{-4.6}$~{\kms} on MJD 60453.27578)\deleted{, validating our detection of the latter}.
However, the RV difference is too large compared to the predicted RV amplitude of PDS 70 b based on orbital solutions in \cite{Wang:2021ab} ($\sim$2.2~{\kms} at the epoch of our observation. We attributed the discrepancy to the systematics and added 2.5~{\kms} as our systematic uncertainty based on the RV analysis in \cite{Ruffio:2023aa}, with our resulting RV = $-1.7^{+3.4}_{-5.2}$~{\kms} for PDS 70 b.
Our {\vsini} = 9$^{+9}_{-7}$~{\kms} ($<$ 29~{\kms} at 95\% confidence level) with a posterior probability distribution peak near 0, so we ran another forward model fit with the {\vsini} set as 0. We found that our result highly favors nondetection of spin in our KPIC spectra of PDS 70b, with $\log_{10} {\mathrm{Bayes \, factor}} = 510.7$ (decisive; \citealp{Jeffreys:1961aa}), and $\Delta \chi^2$ = 779. 
The non-detection of spin is expected as PDS 70b is still accreting and has not yet undergone the contraction and spin-up phase \citep{Bryan:2020ab, Hsu:2024aa}. 
Our KPIC spectra of PDS 70b did not show the Brackett $\gamma$ line emission ($\sim$2.166~{\micron}), similar to previous non-detections reported in \cite{Christiaens:2019aa, Wang:2021ab}.
While our derived {\teff} and {\logg} values are consistent with literature measurements using photometry \citep{Wang:2021ab}, we again stress the limitations of high-resolution spectroscopy in measuring these parameters for low-temperature objects \citep{Hsu:2021aa, Hsu:2024aa}.
To examine whether clouds are necessary to fit the data better, we used the Sonora Bobcat (cloudless) model grids \citep{Marley:2021aa}, and found a worse fit compared to our BT-Settl result ($\log_{10}$ Bayes factor = 0.7, which is substantial following \citealp{Jeffreys:1961aa} and equivalent to 2.4$\sigma$ significance following \citealp{Benneke:2013aa}).

To further validate our detection of PDS 70b if there is indeed planet signal in our data, we ran a stellar model only fit to our KPIC PDS 70b spectra. 
We found that such a star-only model is highly disfavored compared to our combined planet and star model, with $\log_{10} {\mathrm{Bayes \, factor}} = 344.9$, considered decisive \citep{Jeffreys:1961aa}. The $\Delta \chi^2$ of 772 further supports our detection of PDS 70b.

\begin{deluxetable*}{lccc}
\tablecaption{Forward Modeling Priors and Results \label{table:forward_model_parameters}}
\tablecolumns{4}
\tablehead{
\colhead{Description} &  \colhead{Symbol (unit)}  & \colhead{Priors\tablenotemark{a}} & \colhead{Results\tablenotemark{b}}
}
\startdata
\multicolumn{4}{c}{BT-Settl fits} \\
\hline
Effective Temperature & {\teff} (K) & $\mathcal{U}$(800, 1800) & 1103$^{+134}_{-75}$ \\
Surface Gravity & $\log{g}$ (dex cm s$^{-2}$) & $\mathcal{U}$(3.5, 5.5) & 4.7$^{+0.5}_{-0.6}$ \\
Projected Rotational Velocity & {\vsini} ({\kms}) & $\mathcal{U}$(0, 100) & 9$^{+9}_{-7}$\tablenotemark{c} \\
Radial Velocity & RV ({\kms}) & $\mathcal{U}$($-100$, 100) & $-1.7^{+2.3}_{-4.6}$\tablenotemark{d} \\
Planet Flux & Planet Flux (DN) & $\mathcal{U}$(0, 160) & 3.6$^{+1.2}_{-1.1}$ \\
Speckle Flux for Fiber 2\tablenotemark{e} & Speckle Flux (DN) & $\mathcal{U}$($-$50, 200) & 120 $\pm$ 2 \\
Speckle Flux for Fiber 4\tablenotemark{e} & Speckle Flux (DN) & $\mathcal{U}$($-$50, 200) & 110 $\pm$ 2 \\
Scale Term for LSF\tablenotemark{e} & LSF $\sigma$ & $\mathcal{U}$(1.0, 1.2) & 1.11$^{+0.06}_{-0.07}$ \\
Error Jitter\tablenotemark{e} & Error Jitter (DN) & $\mathcal{U}$(0.1, 30.0) & 2.23 $\pm$ 0.02 \\
\hline
\multicolumn{4}{c}{Retrieval fits} \\
\hline
Effective Temperature\tablenotemark{f, g} & {\teff} (K) & 1100 & \nodata \\
Surface Gravity\tablenotemark{f, g} & $\log{g}$ (dex cm s$^{-2}$) & 4.5 & \nodata \\
Projected Rotational Velocity\tablenotemark{g} & {\vsini} ({\kms}) & 0 & \nodata \\
Radial Velocity & RV ({\kms}) & $\mathcal{U}$($-15$, 15) & $-4.4^{+2.0}_{-2.1}$\tablenotemark{d} \\
Carbon/Oxygen & C/O & $\mathcal{U}$(0.1, {1.6}) & {0.28}$^{+0.20}_{-0.12}$ \\
Metallicity & [C/H] & $\mathcal{U}$($-1.5$, 1.5) & ${-0.2}^{+0.8}_{-0.5}$ \\
Quench Pressure & $\log P_\mathrm{quench}$ (bar) & $\mathcal{U}$(${-4}$, ${3}$) & {1.6}$^{+0.9}_{-0.8}$ \\
Planet Flux & Planet Flux (DN) & $\mathcal{U}$(0, 160) & ${2.7}^{+1.2}_{-1.0}$ \\
Speckle Flux for Fiber 2\tablenotemark{e} & Speckle Flux (DN) & $\mathcal{U}$($-$50, 200) & 120 $\pm$ 2 \\
Speckle Flux for Fiber 4\tablenotemark{e} & Speckle Flux (DN) & $\mathcal{U}$($-$50, 200) & 111 $\pm$ 2 \\
Scale Term for LSF\tablenotemark{e} & LSF $\sigma$ & $\mathcal{U}$(1.0, 1.2) & 1.10 $\pm$ 0.07 \\
Error Jitter\tablenotemark{e} & Error Jitter (DN) & $\mathcal{U}$(0.1, 30.0) & 2.22 $\pm$ 0.02 \\
\enddata
\tablenotetext{a}{$\mathcal{U}$ represents uniform priors with the lower and upper bounds in the parentheses.}
\tablenotetext{b}{The uncertainties are reported as the 16$^\mathrm{th}$ and 84$^\mathrm{th}$ percentiles.}
\tablenotetext{c}{Our {\vsini} measurement is consistent with a non-detection  ($<$29~{\kms} at 95\% confidence interval). See Section~\ref{sec:planet_model} for details.}
\tablenotetext{d}{Our reported RV has been corrected with the barycentric velocity during the observation ($-$7.59 {\kms}).}
\tablenotetext{e}{There are three parameters used to fit in orders 31--33. We show only order 33 as an illustration here.}
\tablenotetext{f}{We used the Sonora Bobcat pressure-temperature profile \citep{Marley:2021aa}.}
\tablenotetext{g}{Fixed parameters determined in our BT-Settl results. See Section~\ref{sec:planet_model} and Section~\ref{sec:retrieve} for details.}
\tablenotetext{h}{Our range of C/O priors is limited from 0.1 to 1.6 by the chemistry in \texttt{petitRADTRANS} \citep{Molliere:2020aa}.}
\end{deluxetable*}

\section{Retrieval Analysis} \label{sec:retrieve}

We constrained the C/O ratio and metallicity [C/H] from our KPIC spectra of PDS 70b, using the \texttt{petitRADTRANS} package \citep{Molliere:2019ab, Molliere:2020aa}.
Our method follows the retrieval framework in \cite{Xuan:2022aa} and \cite{Hsu:2024ab}.
Since our KPIC spectra of PDS 70b signals are still relatively noisy (Section~\ref{sec:ccf_detect}), we constrained our retrieval model with parameters determined in the self-consistent BT-Settl model grids in Section~\ref{sec:planet_model}.
We used the Sonora pressure-temperature profile at {\teff} = 1100~K and {\logg} = 4.5~cm s$^{-2}$, and {\vsini} at 0~{\kms} were fixed (Section~\ref{sec:planet_model}) with a cloudless model.
We also restricted our RV priors between $\pm$15~{\kms} to avoid RVs stuck in the unphysical parameter space.
We adopted the chemical disequilibrium model parameterized by the quench pressure $P_\mathrm{quench}$ in \cite{Molliere:2020aa}.
Our full forward retrieval model has 17 parameters in total, including RV, C/O, [C/H], $\log{P_\mathrm{quench}}$, planet flux counts, and 12 nuisance parameters defined in Section~\ref{sec:planet_model}.

For the molecular species, we used the line-by-line approach, including all CO isotopologues (`CO\_all\_iso'), the principal H$_2$O isotopologue (`H2O\_main\_iso') from the HITEMP database \citep{Rothman:2010aa}, and the parent CH$_4$ species (`CH4\_hargreaves\_main\_iso') from \cite{Hargreaves:2020aa}. 
We also included Rayleigh scattering from hydrogen and helium \citep{Dalgarno:1962aa, Chan:1965aa} and continuum collision-induced absorption (CIA) opacities for H$_2$-H$_2$ and H$_2$-He \citep{Gray:2008aa}.
To expedite the computation, the resolution of the opacities ($R = 10^6$) was downsampled by a factor of 4 (i.e. to $R = 2.5 \times 10^5$).

We again used the nested sampling method with 1000 live points (as in Section~\ref{sec:planet_model}) to derive our best-fit model.
The fixed parameters, priors, and best-fit parameters are summarized in Table~\ref{table:forward_model_parameters}.
Figure~\ref{fig:btsettl_posterior} shows the posteriors for RV, C/O and [C/H] from our best-fit forward retrieval model.
Our best-fit values are C/O = 0.28$^{+0.20}_{-0.12}$ ($\lesssim$ {0.63} at 95\% confidence level) and [C/H] = ${-0.2}^{+0.8}_{-0.5}$~dex, which are consistent with solar and the host star metallicity.
Our best-fit RV = $-4.4^{+2.0}_{-2.1}$~{\kms} is consistent with our RV derived in our BT-Settl fit.



\section{Discussion} \label{sec:discuss}

One route to constraining the formation history of PDS 70b is to 
place our measured C/O ratio and metallicty for the protoplanet (C/O $\lesssim$ 0.63; Section~\ref{sec:retrieve}) in the context of that of the host star PDS 70 A (C/O $\sim$ 0.44; \citealp{Cridland:2023aa}) and the outer gas disk (C/O $\gtrsim 1$; \citealp{Facchini:2021aa, Cridland:2023aa, Law:2024aa}).
We note that the ALMA gas disk C/O measurements in \cite{Facchini:2021aa, Cridland:2023aa, Law:2024aa} only probe the gas phase C and O reservoirs, typically at large scale heights in the disk, whereas the solids are often located near the disk midplane, and with compositions that are impossible to constrain with the current facilities.
Were the metallicity of PDS 70b substellar but with a high C/O ratio, this would likely indicate that the bulk of the accretion onto PDS 70b came from the gas phase of the disk within its Hill radius \citep{Oberg:2011ab}.
However, our measured C/O ({0.28}$^{+0.20}_{-0.12}$) is consistent with the host star ($0.44 \pm 0.19$; \citealp{Cridland:2023aa}), and we find $\sim$stellar metallicity. The high outer gas disk C/O value of $\gtrsim 1$ measured by ALMA \citep{Facchini:2021aa, Cridland:2023aa} and the PDS 70b values suggest that only a small fraction of carbon in the gas phase was and is within PDS 70b's Hill radius, and that most of the carbon incorporated into PDS 70b arrived via ice and dust aggregates/solids.

In scenarios where the gas-phase carbon and oxygen are highly depleted in the outer disk, a small amount of accretion of solids onto PDS 70b can overwhelm the C/O contribution from the gas.
Such gas-phase depletion, especially of CO, has been reported in other protoplanetary disks (e.g., TW Hya, DM Tau, IM Lup), and can be explained by chemical reactions or physical processes such as vertical mixing, freezing-out onto dust grain, and radial drift. Thus, such depletion can be both radially dependent and time variable \citep{Favre:2013aa, Zhang:2019ac, Yoshida:2022aa}, as exemplified by disks that do not exhibit CO depletion such as HD 163296 \citep{Zhang:2019ac} or many of the objects studied in an ALMA disk survey of the Lupus star-forming region \citep{Pascucci:2023aa}. 

A more accurate interpretation of the C/O in exoplanet atmospheres requires modeling of disk evolution, specifically the potentially divergent paths followed by dust/ice versus gas.
One such model, by \cite{Cridland:2023aa}, provides an interpretation of our measured C/O and metallicity patterns in which the stellar-like C/O of PDS 70b is explained by late stage carbon-enrichment of the disk gas {\em after} the planet was formed.
This model tracks the core accretion of volatile and refractory C, O species accretions onto the growing PDS 70b and c planets, and simulates the concurrent evolution of the gas-phase chemistry in the disk, focusing on $^{12}$CO, C$^{18}$O, and C$_2$H to track the C/O evolution.
The authors find that the C/O values of the planets depend on the time when the disk was enriched with carbon through chemical or physical processes.
If carbon is enriched throughout the disk lifetime, the planets would present a superstellar C/O ratio.
Alternatively, if the disk carbon enrichment occurred relatively recently -- after the planet was formed -- PDS 70b will instead show a C/O value similar to its host star, consistent with our measurements.
While not considered in the model above, photochemistry that sublimes and thus removes carbon in the pebbles during the planet's runaway gas accretion could also contribute to the observed lower C/O of the planet and the local C/O enhancement in the gas disk \citep{Jiang:2023aa}.

Growing evidence for the substantial evolution of the volatile element chemistry in the warm, inner region of disks has come from {\em JWST}. \citep{Henning24, Pontoppidan24}.
Thought to be driven by divergent dust+ice versus gas evolution, and first hinted at by {\em Spizter}, {\em JWST}/MIRI spectra in particular reveal disk gas evolution from O-enriched at young (up to few Myr) ages \citep{Banzatti23,Banzatti24,Munoz-Romero24} into a carbon-rich phase \citep{Pascucci13,Tabone23,Arabhavi:2024sc}, especially in the disks around very low-mass stars. Recently, disks in possible transition between these phases have been found around Sun-like stars with low accretion rates, that display rich mixtures of water and complex carbonaceous species \citep{Colmenares24}. 

Placing the true nature of the formation history of PDS 70b into this general context will require more sophisticated disk modeling, ideally as a function of radius, along with our PDS 70b C/O measurement, to assess if the C/O of the gas at the location of PDS 70b is the same as the global C/O of the gas disk. 
In addition, the accretion history of PDS 70b, including the C/O of the dust and dust-to-gas ratio, should be considered more carefully in future modeling studies.

Observationally, we emphasize that our inferred parameters are based on our relatively noisy KPIC spectra (CO + H$_2$O CCF SNR $\sim$ 4.2). Future observations of KPIC, VLT/CRIRES+ \citep{Follert:2014aa}, VLT/HiRISE \citep{Otten:2021aa}, Subaru/REACH \citep{Kotani:2020aa}, HISPEC \citep{Mawet:2019aa, Konopacky:2023aa}, and \textit{JWST}/MIRI \citep{Wells:2015aa} will verify and improve our best estimates of the physical parameters of PDS 70b and possibly PDS 70c.
While higher SNR measurements of C/O ratios will still be challenging to distinguish these two formation scenarios, the isotope ratios can be constrained to infer their accretion history of carbon and oxygen in the disk midplane, for example, $^{12}$C/$^{13}$C and $^{16}$O/$^{18}$O \citep{Zhang:2021ae, Xuan:2024aa, Xuan:2024ab}, which requires very high CCF SNR $>$ 50 for CO and H$_2$O main isotope species. 
Another way to disentangle the two scenarios from atmospheric characterizations of the planets is by measuring the refractory species such as Fe and Si in the near- (FeH in \textit{J}- and \textit{H}-band; \citealp{Cushing:2005aa}) and mid-infrared (silicate clouds; \citealp{Suarez:2022aa}) \citep{Chachan:2023aa}. 
On the other hand, a more accurate inference of PDS 70 planet formation history requires significant modeling work including tracing the volatile and refractory species, and the associated isotopes for the planet and disk evolution in the future.
Future discoveries and characterizations of more protoplanets for trends in the population of protoplanet compositions, with the next-generation telescopes such as the Giant Magellan Telescope, Thirty Meter Telescope, and Extremely Large Telescope will allow us to statistically understand the formation and evolution of gas giant exoplanets at 10s of au from their host stars.

\section{Summary}\label{sec:sum}

In this Letter, we reported the detection and characterization of the superjovian protoplanet PDS 70b using Keck/KPIC high-resolution spectroscopy, with detections of atmospheric CO and H$_2$O.

Our detection of PDS 70b relies on a least-squares-based cross-correlation function (CCF), with the CCF SNRs for CO at 3.8~$\sigma$, H$_2$O at 3.5~$\sigma$, and CO and H$_2$O template at 4.2~$\sigma$.
We fit our KPIC spectra and derived a barycentric corrected radial velocity of PDS 70b (RV = $-1.7^{+3.4}_{-5.2}$~{\kms}) is different from its host star during the same epoch (RV = $+$6.65$^{+0.14}_{-0.22}$~{\kms}), which is \replaced{3.6}{2.5}~$\sigma$s difference\deleted{, supporting our detection of the planet in our KPIC high-resolution spectra}.
We found the PDS 70b is consistent with non-detection of spin ($v\sin{i}$ $<$ 29~{\kms} at 95\% confidence level), expected given its ongoing accretion and contraction, and its large radius ($2.7^{+0.4}_{-0.3}$~R$_\mathrm{Jup}$; \citealp{Wang:2020aa}). 
We further validated our detection by comparing the star-only fit with our star and planet model and found that the star-only fit is highly disfavored by $\chi^2$ ($\Delta \chi^2$ = 772) and the Bayes factor ($\log_{10}$ Bayes factor = 344.9) statistics.
Our atmospheric retrievals constrained the C/O = {0.28}$^{+0.20}_{-0.12}$ ($\lesssim$ {0.63} at 95\% confidence level) and [C/H] = ${-0.2}^{+0.8}_{-0.5}$~dex for PDS 70b.

Our measured C/O and metallicity of PDS 70b are consistent with those of the host star PDS 70 A (CO $\sim$ 0.44; [Fe/H] = $-0.11 \pm 0.19$~dex), and lower than the gas-phase C/O of PDS 70 outer disk ($\gtrsim$ 1).
The stellar-like C/O of PDS 70b can be interpreted via two scenarios.
PDS 70b might have accreted the bulk materials of its carbon and oxygen from solids (dust+ice), instead of gas phase volatiles, or PDS 70b might have formed before the disk gas was enriched in carbon.
These scenarios cannot be distinguished by the existing the C/O measurements of the star, planet, and its disk, due to missing information on the accretion history of the planet PDS 70b, the dust C/O ratio, and the time dependent dust-to-gas ratio.
Future observations, with higher SNRs of PDS 70b and possibly PDS 70c for both volatile and refractory species are required to refine and validate the true chemistry in their atmospheres, as well as a more detailed modeling of the whole system.

\facilities{Keck:II (KPIC), Keck:II (NIRSPEC), Keck:II (NIRC2)}

\software{
\texttt{Astropy} \citep{Astropy-Collaboration:2013aa,Astropy-Collaboration:2018aa}, 
\texttt{DYNESTY} \citep{Speagle:2020aa}, 
\texttt{corner} \citep{Foreman-Mackey:2016aa},
\texttt{emcee} \citep{Foreman-Mackey:2013aa},
\texttt{Matplotlib} \citep{Hunter:2007aa}, 
\texttt{Numpy} \citep{Harris:2020aa},
\texttt{petitRADTRANS} \citep{Molliere:2019ab},
\texttt{Scipy} \citep{Virtanen:2020aa}, 
\texttt{seaborn} \citep{Waskom2021}, 
\texttt{SMART} \citep{Hsu:2021aa, Hsu:2021ab}
          }


\begin{acknowledgments}
The authors thank Yayaati Chachan, Stefano Facchini, and Bruce Macintosh for extremely useful discussions on the physical interpretation of planet and disk chemistry, which have significantly improved this manuscript.
The authors thank the anonymous referee for useful comments, which improved the original manuscript.
The authors thank the Keck observing assistant Arina Rostopchina for her help in obtaining the Keck/KPIC spectra.
W. M. Keck Observatory access was supported by Northwestern University and the Center for Interdisciplinary Exploration and Research in Astrophysics (CIERA).
This research was supported in part through the computational resources and staff contributions provided for the Quest high-performance computing facility at Northwestern University which is jointly supported by the Office of the Provost, the Office for Research, and Northwestern University Information Technology.
This work used computing resources provided by Northwestern University and the Center for Interdisciplinary Exploration and Research in Astrophysics (CIERA).
Funding for KPIC has been provided by the California Institute of Technology, the Jet Propulsion Laboratory, the Heising-Simons Foundation (grants \#2015-129, \#2017-318, \#2019-1312, and \#2023-4598), the Simons Foundation (through the Caltech Center for Comparative Planetary Evolution), and the NSF under grant AST-1611623.
This research has made use of the NASA Exoplanet Archive, which is operated by the California Institute of Technology, under contract with the National Aeronautics and Space Administration under the Exoplanet Exploration Program.
Data presented herein were obtained at the W. M. Keck Observatory, which is operated as a scientific partnership among the California Institute of Technology, the University of California, and the National Aeronautics and Space Administration. The Observatory was made possible by the generous financial support of the W. M. Keck Foundation. 
The authors recognize and acknowledge the significant cultural role
and reverence that the summit of Maunakea has with the
indigenous Hawaiian community, and that the W. M. Keck
Observatory stands on Crown and Government Lands that the
State of Hawai'i is obligated to protect and preserve for future
generations of indigenous Hawaiians. 

\end{acknowledgments}

\clearpage

\bibliography{main}{}

\begin{thebibliography}{}
\expandafter\ifx\csname natexlab\endcsname\relax\def\natexlab#1{#1}\fi
\providecommand{\url}[1]{\href{#1}{#1}}
\providecommand{\dodoi}[1]{doi:~\href{http://doi.org/#1}{\nolinkurl{#1}}}
\providecommand{\doeprint}[1]{\href{http://ascl.net/#1}{\nolinkurl{http://ascl.net/#1}}}
\providecommand{\doarXiv}[1]{\href{https://arxiv.org/abs/#1}{\nolinkurl{https://arxiv.org/abs/#1}}}

\bibitem[{Arabhavi {et~al.}(2024)Arabhavi, Kamp, Henning, van Dishoeck, Christiaens, Gasman, Perrin, G{\"u}del, Tabone, Kanwar, Waters, Pascucci, Samland, Perotti, Bettoni, Grant, Lagage, Ray, Vandenbussche, Absil, Argyriou, Barrado, Boccaletti, Bouwman, o~Garatti, Glauser, Lahuis, Mueller, Olofsson, Pantin, Scheithauer, Morales-Calder{\'o}n, Franceschi, Jang, Pawellek, Rodgers-Lee, Schreiber, Schwarz, Temmink, Vlasblom, Wright, Colina, \& {\"O}stlin}]{Arabhavi:2024sc}
Arabhavi, A.~M., Kamp, I., Henning, T., {et~al.} 2024, Science, 384, 1086, \dodoi{10.1126/science.adi8147}

\bibitem[{{Astropy Collaboration} {et~al.}(2013){Astropy Collaboration}, {Robitaille}, {Tollerud}, {Greenfield}, {Droettboom}, {Bray}, {Aldcroft}, {Davis}, {Ginsburg}, {Price-Whelan}, {Kerzendorf}, {Conley}, {Crighton}, {Barbary}, {Muna}, {Ferguson}, {Grollier}, {Parikh}, {Nair}, {Unther}, {Deil}, {Woillez}, {Conseil}, {Kramer}, {Turner}, {Singer}, {Fox}, {Weaver}, {Zabalza}, {Edwards}, {Azalee Bostroem}, {Burke}, {Casey}, {Crawford}, {Dencheva}, {Ely}, {Jenness}, {Labrie}, {Lim}, {Pierfederici}, {Pontzen}, {Ptak}, {Refsdal}, {Servillat}, \& {Streicher}}]{Astropy-Collaboration:2013aa}
{Astropy Collaboration}, {Robitaille}, T.~P., {Tollerud}, E.~J., {et~al.} 2013, \aap, 558, A33, \dodoi{10.1051/0004-6361/201322068}

\bibitem[{{Astropy Collaboration} {et~al.}(2018){Astropy Collaboration}, {Price-Whelan}, {Sip{\H{o}}cz}, {G{\"u}nther}, {Lim}, {Crawford}, {Conseil}, {Shupe}, {Craig}, {Dencheva}, {Ginsburg}, {VanderPlas}, {Bradley}, {P{\'e}rez-Su{\'a}rez}, {de Val-Borro}, {Aldcroft}, {Cruz}, {Robitaille}, {Tollerud}, {Ardelean}, {Babej}, {Bach}, {Bachetti}, {Bakanov}, {Bamford}, {Barentsen}, {Barmby}, {Baumbach}, {Berry}, {Biscani}, {Boquien}, {Bostroem}, {Bouma}, {Brammer}, {Bray}, {Breytenbach}, {Buddelmeijer}, {Burke}, {Calderone}, {Cano Rodr{\'\i}guez}, {Cara}, {Cardoso}, {Cheedella}, {Copin}, {Corrales}, {Crichton}, {D'Avella}, {Deil}, {Depagne}, {Dietrich}, {Donath}, {Droettboom}, {Earl}, {Erben}, {Fabbro}, {Ferreira}, {Finethy}, {Fox}, {Garrison}, {Gibbons}, {Goldstein}, {Gommers}, {Greco}, {Greenfield}, {Groener}, {Grollier}, {Hagen}, {Hirst}, {Homeier}, {Horton}, {Hosseinzadeh}, {Hu}, {Hunkeler}, {Ivezi{\'c}}, {Jain}, {Jenness}, {Kanarek}, {Kendrew}, {Kern}, {Kerzendorf}, {Khvalko}, {King}, {Kirkby}, {Kulkarni},
  {Kumar}, {Lee}, {Lenz}, {Littlefair}, {Ma}, {Macleod}, {Mastropietro}, {McCully}, {Montagnac}, {Morris}, {Mueller}, {Mumford}, {Muna}, {Murphy}, {Nelson}, {Nguyen}, {Ninan}, {N{\"o}the}, {Ogaz}, {Oh}, {Parejko}, {Parley}, {Pascual}, {Patil}, {Patil}, {Plunkett}, {Prochaska}, {Rastogi}, {Reddy Janga}, {Sabater}, {Sakurikar}, {Seifert}, {Sherbert}, {Sherwood-Taylor}, {Shih}, {Sick}, {Silbiger}, {Singanamalla}, {Singer}, {Sladen}, {Sooley}, {Sornarajah}, {Streicher}, {Teuben}, {Thomas}, {Tremblay}, {Turner}, {Terr{\'o}n}, {van Kerkwijk}, {de la Vega}, {Watkins}, {Weaver}, {Whitmore}, {Woillez}, {Zabalza}, \& {Astropy Contributors}}]{Astropy-Collaboration:2018aa}
{Astropy Collaboration}, {Price-Whelan}, A.~M., {Sip{\H{o}}cz}, B.~M., {et~al.} 2018, \aj, 156, 123, \dodoi{10.3847/1538-3881/aabc4f}

\bibitem[{{Banzatti} {et~al.}(2023){Banzatti}, {Pontoppidan}, {Carr}, {Jellison}, {Pascucci}, {Najita}, {Romero-Mirza}, {{\"O}berg}, {Kalyaan}, {Pinilla}, {Krijt}, {Long}, {Lambrechts}, {Rosotti}, {Herczeg}, {Salyk}, {Zhang}, {Bergin}, {Ballering}, {Meyer}, {Bruderer}, \& {Jdiscs Collaboration}}]{Banzatti23}
{Banzatti}, A., {Pontoppidan}, K.~M., {Carr}, J.~S., {et~al.} 2023, \apjl, 957, L22, \dodoi{10.3847/2041-8213/acf5ec}

\bibitem[{{Banzatti} {et~al.}(2024){Banzatti}, {Salyk}, {Pontoppidan}, {Carr}, {Zhang}, {Arulanantham}, {Cleeves}, {Krijt}, {Najita}, {Oberg}, {Pascucci}, {Blake}, {Munoz-Romero}, {Bergin}, {Cieza}, {Pinilla}, {Long}, {Mallaney}, {Xie}, \& {the JDISCS collaboration}}]{Banzatti24}
{Banzatti}, A., {Salyk}, C., {Pontoppidan}, K.~M., {et~al.} 2024, arXiv e-prints, arXiv:2409.16255, \dodoi{10.48550/arXiv.2409.16255}

\bibitem[{{Baraffe} {et~al.}(2015){Baraffe}, {Homeier}, {Allard}, \& {Chabrier}}]{Baraffe:2015aa}
{Baraffe}, I., {Homeier}, D., {Allard}, F., \& {Chabrier}, G. 2015, \aap, 577, A42, \dodoi{10.1051/0004-6361/201425481}

\bibitem[{{Benneke} \& {Seager}(2013)}]{Benneke:2013aa}
{Benneke}, B., \& {Seager}, S. 2013, \apj, 778, 153, \dodoi{10.1088/0004-637X/778/2/153}

\bibitem[{{Bowler}(2016)}]{Bowler:2016aa}
{Bowler}, B.~P. 2016, \pasp, 128, 102001, \dodoi{10.1088/1538-3873/128/968/102001}

\bibitem[{{Bowler} {et~al.}(2023){Bowler}, {Tran}, {Zhang}, {Morgan}, {Ashok}, {Blunt}, {Bryan}, {Evans}, {Franson}, {Huber}, {Nagpal}, {Wu}, \& {Zhou}}]{Bowler:2023aa}
{Bowler}, B.~P., {Tran}, Q.~H., {Zhang}, Z., {et~al.} 2023, \aj, 165, 164, \dodoi{10.3847/1538-3881/acbd34}

\bibitem[{{Bryan} {et~al.}(2020){Bryan}, {Ginzburg}, {Chiang}, {Morley}, {Bowler}, {Xuan}, \& {Knutson}}]{Bryan:2020ab}
{Bryan}, M.~L., {Ginzburg}, S., {Chiang}, E., {et~al.} 2020, \apj, 905, 37, \dodoi{10.3847/1538-4357/abc0ef}

\bibitem[{{Cale} {et~al.}(2019){Cale}, {Plavchan}, {LeBrun}, {Gagn{\'e}}, {Gao}, {Tanner}, {Beichman}, {Xuesong Wang}, {Gaidos}, {Teske}, {Ciardi}, {Vasisht}, {Kane}, \& {von Braun}}]{Cale:2019aa}
{Cale}, B., {Plavchan}, P., {LeBrun}, D., {et~al.} 2019, \aj, 158, 170, \dodoi{10.3847/1538-3881/ab3b0f}

\bibitem[{{Chachan} {et~al.}(2023){Chachan}, {Knutson}, {Lothringer}, \& {Blake}}]{Chachan:2023aa}
{Chachan}, Y., {Knutson}, H.~A., {Lothringer}, J., \& {Blake}, G.~A. 2023, \apj, 943, 112, \dodoi{10.3847/1538-4357/aca614}

\bibitem[{{Chan} \& {Dalgarno}(1965)}]{Chan:1965aa}
{Chan}, Y.~M., \& {Dalgarno}, A. 1965, Proceedings of the Physical Society, 85, 227, \dodoi{10.1088/0370-1328/85/2/304}

\bibitem[{{Christiaens} {et~al.}(2019){Christiaens}, {Casassus}, {Absil}, {Cantalloube}, {Gomez Gonzalez}, {Girard}, {Ram{\'\i}rez}, {Pairet}, {Salinas}, {Price}, {Pinte}, {Quanz}, {Jord{\'a}n}, {Mawet}, \& {Wahhaj}}]{Christiaens:2019aa}
{Christiaens}, V., {Casassus}, S., {Absil}, O., {et~al.} 2019, \mnras, 486, 5819, \dodoi{10.1093/mnras/stz1232}

\bibitem[{Colmenares {et~al.}(2024)Colmenares, Bergin, Salyk, Pontopiddan, Arulanantham, Calahan, Banzatti, Andrews, Blake, Ciesla, Green, Long, Lambrechts, Najita, Pascucci, Pinilla, Krijt, Trapman, \& the JDISCS~Collaboration}]{Colmenares24}
Colmenares, M.~J., Bergin, E., Salyk, C., {et~al.} 2024, JWST/MIRI detection of a carbon-rich chemistry in a solar nebula analog.
\newblock \doarXiv{2410.18187}

\bibitem[{{Cowley} {et~al.}(1969){Cowley}, {Cowley}, {Jaschek}, \& {Jaschek}}]{Cowley:1969aa}
{Cowley}, A., {Cowley}, C., {Jaschek}, M., \& {Jaschek}, C. 1969, \aj, 74, 375, \dodoi{10.1086/110819}

\bibitem[{{Cridland} {et~al.}(2023){Cridland}, {Facchini}, {van Dishoeck}, \& {Benisty}}]{Cridland:2023aa}
{Cridland}, A.~J., {Facchini}, S., {van Dishoeck}, E.~F., \& {Benisty}, M. 2023, \aap, 674, A211, \dodoi{10.1051/0004-6361/202245619}

\bibitem[{{Cugno} {et~al.}(2021){Cugno}, {Patapis}, {Stolker}, {Quanz}, {Boehle}, {Hoeijmakers}, {Marleau}, {Molli{\`e}re}, {Nasedkin}, \& {Snellen}}]{Cugno:2021aa}
{Cugno}, G., {Patapis}, P., {Stolker}, T., {et~al.} 2021, \aap, 653, A12, \dodoi{10.1051/0004-6361/202140632}

\bibitem[{{Currie} {et~al.}(2023){Currie}, {Biller}, {Lagrange}, {Marois}, {Guyon}, {Nielsen}, {Bonnefoy}, \& {De Rosa}}]{Currie:2023ab}
{Currie}, T., {Biller}, B., {Lagrange}, A., {et~al.} 2023, in Astronomical Society of the Pacific Conference Series, Vol. 534, Protostars and Planets VII, ed. S.~{Inutsuka}, Y.~{Aikawa}, T.~{Muto}, K.~{Tomida}, \& M.~{Tamura}, 799, \dodoi{10.26624/UCTD5384}

\bibitem[{{Currie} {et~al.}(2022){Currie}, {Lawson}, {Schneider}, {Lyra}, {Wisniewski}, {Grady}, {Guyon}, {Tamura}, {Kotani}, {Kawahara}, {Brandt}, {Uyama}, {Muto}, {Dong}, {Kudo}, {Hashimoto}, {Fukagawa}, {Wagner}, {Lozi}, {Chilcote}, {Tobin}, {Groff}, {Ward-Duong}, {Januszewski}, {Norris}, {Tuthill}, {van der Marel}, {Sitko}, {Deo}, {Vievard}, {Jovanovic}, {Martinache}, \& {Skaf}}]{Currie:2022aa}
{Currie}, T., {Lawson}, K., {Schneider}, G., {et~al.} 2022, Nature Astronomy, 6, 751, \dodoi{10.1038/s41550-022-01634-x}

\bibitem[{{Cushing} {et~al.}(2005){Cushing}, {Rayner}, \& {Vacca}}]{Cushing:2005aa}
{Cushing}, M.~C., {Rayner}, J.~T., \& {Vacca}, W.~D. 2005, \apj, 623, 1115, \dodoi{10.1086/428040}

\bibitem[{{Cutri} {et~al.}(2003){Cutri}, {Skrutskie}, {van Dyk}, {Beichman}, {Carpenter}, {Chester}, {Cambresy}, {Evans}, {Fowler}, {Gizis}, {Howard}, {Huchra}, {Jarrett}, {Kopan}, {Kirkpatrick}, {Light}, {Marsh}, {McCallon}, {Schneider}, {Stiening}, {Sykes}, {Weinberg}, {Wheaton}, {Wheelock}, \& {Zacarias}}]{Cutri:2003aa}
{Cutri}, R.~M., {Skrutskie}, M.~F., {van Dyk}, S., {et~al.} 2003, {2MASS All Sky Catalog of point sources.}

\bibitem[{{Dalgarno} \& {Williams}(1962)}]{Dalgarno:1962aa}
{Dalgarno}, A., \& {Williams}, D.~A. 1962, \apj, 136, 690, \dodoi{10.1086/147428}

\bibitem[{{Delorme} {et~al.}(2021){Delorme}, {Jovanovic}, {Echeverri}, {Mawet}, {Kent Wallace}, {Bartos}, {Cetre}, {Wizinowich}, {Ragland}, {Lilley}, {Wetherell}, {Doppmann}, {Wang}, {Morris}, {Ruffio}, {Martin}, {Fitzgerald}, {Ruane}, {Schofield}, {Suominen}, {Calvin}, {Wang}, {Magnone}, {Johnson}, {Sohn}, {L{\'o}pez}, {Bond}, {Pezzato}, {Sayson}, {Chun}, \& {Skemer}}]{Delorme:2021aa}
{Delorme}, J.-R., {Jovanovic}, N., {Echeverri}, D., {et~al.} 2021, Journal of Astronomical Telescopes, Instruments, and Systems, 7, 035006, \dodoi{10.1117/1.JATIS.7.3.035006}

\bibitem[{Echeverri {et~al.}(2024)Echeverri, Jovanovic, Delorme, Guthery, Roberts, Nash, Horstman, Xuan, Xin, Finnerty, Hsu, Ruane, Leifer, Zimmer, Soda, Schofield, Wallace, Wang, Mawet, Marin, Wizinowich, Ge, \& Gao}]{Echeverri:2024aa}
Echeverri, D., Jovanovic, N., Delorme, J.-R., {et~al.} 2024, in Ground-based and Airborne Instrumentation for Astronomy X, ed. J.~J. Bryant, K.~Motohara, \& J.~R.~D. Vernet, Vol. 13096, International Society for Optics and Photonics (SPIE), 130962D, \dodoi{10.1117/12.3019085}

\bibitem[{{Facchini} {et~al.}(2021){Facchini}, {Teague}, {Bae}, {Benisty}, {Keppler}, \& {Isella}}]{Facchini:2021aa}
{Facchini}, S., {Teague}, R., {Bae}, J., {et~al.} 2021, \aj, 162, 99, \dodoi{10.3847/1538-3881/abf0a4}

\bibitem[{{Favre} {et~al.}(2013){Favre}, {Cleeves}, {Bergin}, {Qi}, \& {Blake}}]{Favre:2013aa}
{Favre}, C., {Cleeves}, L.~I., {Bergin}, E.~A., {Qi}, C., \& {Blake}, G.~A. 2013, \apjl, 776, L38, \dodoi{10.1088/2041-8205/776/2/L38}

\bibitem[{{Follert} {et~al.}(2014){Follert}, {Dorn}, {Oliva}, {Lizon}, {Hatzes}, {Piskunov}, {Reiners}, {Seemann}, {Stempels}, {Heiter}, {Marquart}, {Lockhart}, {Anglada-Escude}, {L{\"o}winger}, {Baade}, {Grunhut}, {Bristow}, {Klein}, {Jung}, {Ives}, {Kerber}, {Pozna}, {Paufique}, {Kaeufl}, {Origlia}, {Valenti}, {Gojak}, {Hilker}, {Pasquini}, {Smette}, \& {Smoker}}]{Follert:2014aa}
{Follert}, R., {Dorn}, R.~J., {Oliva}, E., {et~al.} 2014, in Society of Photo-Optical Instrumentation Engineers (SPIE) Conference Series, Vol. 9147, Ground-based and Airborne Instrumentation for Astronomy V, ed. S.~K. {Ramsay}, I.~S. {McLean}, \& H.~{Takami}, 914719, \dodoi{10.1117/12.2054197}

\bibitem[{Foreman-Mackey(2016)}]{Foreman-Mackey:2016aa}
Foreman-Mackey, D. 2016, The Journal of Open Source Software, 1, 24, \dodoi{10.21105/joss.00024}

\bibitem[{{Foreman-Mackey} {et~al.}(2013){Foreman-Mackey}, {Hogg}, {Lang}, \& {Goodman}}]{Foreman-Mackey:2013aa}
{Foreman-Mackey}, D., {Hogg}, D.~W., {Lang}, D., \& {Goodman}, J. 2013, \pasp, 125, 306, \dodoi{10.1086/670067}

\bibitem[{{Gaia Collaboration} {et~al.}(2018){Gaia Collaboration}, {Brown}, {Vallenari}, {Prusti}, {de Bruijne}, {Babusiaux}, {Bailer-Jones}, {Biermann}, {Evans}, {Eyer}, \& et~al.}]{Gaia-Collaboration:2018ab}
{Gaia Collaboration}, {Brown}, A.~G.~A., {Vallenari}, A., {et~al.} 2018, \aap, 616, A1, \dodoi{10.1051/0004-6361/201833051}

\bibitem[{{Gaia Collaboration} {et~al.}(2023){Gaia Collaboration}, {Vallenari}, {Brown}, {Prusti}, {de Bruijne}, {Arenou}, {Babusiaux}, {Biermann}, {Creevey}, {Ducourant}, \& et~al.}]{Gaia-Collaboration:2023aa}
{Gaia Collaboration}, {Vallenari}, A., {Brown}, A.~G.~A., {et~al.} 2023, \aap, 674, A1, \dodoi{10.1051/0004-6361/202243940}

\bibitem[{{Gray}(2008)}]{Gray:2008aa}
{Gray}, D.~F. 2008, {The Observation and Analysis of Stellar Photospheres}

\bibitem[{{Haffert} {et~al.}(2019){Haffert}, {Bohn}, {de Boer}, {Snellen}, {Brinchmann}, {Girard}, {Keller}, \& {Bacon}}]{Haffert:2019aa}
{Haffert}, S.~Y., {Bohn}, A.~J., {de Boer}, J., {et~al.} 2019, Nature Astronomy, 3, 749, \dodoi{10.1038/s41550-019-0780-5}

\bibitem[{{Hargreaves} {et~al.}(2020){Hargreaves}, {Gordon}, {Rey}, {Nikitin}, {Tyuterev}, {Kochanov}, \& {Rothman}}]{Hargreaves:2020aa}
{Hargreaves}, R.~J., {Gordon}, I.~E., {Rey}, M., {et~al.} 2020, \apjs, 247, 55, \dodoi{10.3847/1538-4365/ab7a1a}

\bibitem[{Harris {et~al.}(2020)Harris, Millman, van~der Walt, Gommers, Virtanen, Cournapeau, Wieser, Taylor, Berg, Smith, Kern, Picus, Hoyer, van Kerkwijk, Brett, Haldane, del R{\'{i}}o, Wiebe, Peterson, G{\'{e}}rard-Marchant, Sheppard, Reddy, Weckesser, Abbasi, Gohlke, \& Oliphant}]{Harris:2020aa}
Harris, C.~R., Millman, K.~J., van~der Walt, S.~J., {et~al.} 2020, Nature, 585, 357, \dodoi{10.1038/s41586-020-2649-2}

\bibitem[{{Hashimoto} {et~al.}(2020){Hashimoto}, {Aoyama}, {Konishi}, {Uyama}, {Takasao}, {Ikoma}, \& {Tanigawa}}]{Hashimoto:2020aa}
{Hashimoto}, J., {Aoyama}, Y., {Konishi}, M., {et~al.} 2020, \aj, 159, 222, \dodoi{10.3847/1538-3881/ab811e}

\bibitem[{{Henning} {et~al.}(2024){Henning}, {Kamp}, {Samland}, {Arabhavi}, {Kanwar}, {van Dishoeck}, {G{\"u}del}, {Lagage}, {Waelkens}, {Abergel}, {Absil}, {Barrado}, {Boccaletti}, {Bouwman}, {Caratti o Garatti}, {Geers}, {Glauser}, {Lahuis}, {Mueller}, {Nehm{\'e}}, {Olofsson}, {Pantin}, {Ray}, {Scheithauer}, {Vandenbussche}, {Waters}, {Wright}, {Argyriou}, {Christiaens}, {Franceschi}, {Gasman}, {Grant}, {Guadarrama}, {Jang}, {Morales-Calder{\'o}n}, {Pawellek}, {Perotti}, {Rodgers-Lee}, {Schreiber}, {Schwarz}, {Tabone}, {Temmink}, {Vlasblom}, {Colina}, {Greve}, \& {{\"O}stlin}}]{Henning24}
{Henning}, T., {Kamp}, I., {Samland}, M., {et~al.} 2024, \pasp, 136, 054302, \dodoi{10.1088/1538-3873/ad3455}

\bibitem[{{Horne}(1986)}]{Horne:1986aa}
{Horne}, K. 1986, \pasp, 98, 609, \dodoi{10.1086/131801}

\bibitem[{Horstman {et~al.}(2024)Horstman, Ruffio, Wang, Hsu, Baker, Finnerty, Xuan, Echeverri, Mawet, Blake, Bartos, Bond, Calvin, Cetre, Delorme, Doppmann, Fitzgerald, Jovanovic, Lopez, Martin, Morris, Pezzato, Ruane, Sappey, Schofield, Skemer, Venenciano, Wallace, Wang, \& Wizinowich}]{Horstman:2024aa}
Horstman, K.~A., Ruffio, J.-B., Wang, J.~J., {et~al.} 2024, in Ground-based and Airborne Instrumentation for Astronomy X, ed. J.~J. Bryant, K.~Motohara, \& J.~R.~D. Vernet, Vol. 13096, International Society for Optics and Photonics (SPIE), 130962E, \dodoi{10.1117/12.3018020}

\bibitem[{{Hsu} {et~al.}(2023){Hsu}, {Burgasser}, \& {Theissen}}]{Hsu:2023aa}
{Hsu}, C.-C., {Burgasser}, A.~J., \& {Theissen}, C.~A. 2023, \apjl, 945, L6, \dodoi{10.3847/2041-8213/acba8c}

\bibitem[{Hsu {et~al.}(2021)Hsu, Theissen, Burgasser, \& Birky}]{Hsu:2021ab}
Hsu, C.-C., Theissen, C., Burgasser, A., \& Birky, J. 2021, SMART: The Spectral Modeling Analysis and RV Tool, v1.0.0,  Zenodo, \dodoi{10.5281/zenodo.4765258}

\bibitem[{{Hsu} {et~al.}(2021){Hsu}, {Burgasser}, {Theissen}, {Gelino}, {Birky}, {Diamant}, {Bardalez Gagliuffi}, {Aganze}, {Blake}, \& {Faherty}}]{Hsu:2021aa}
{Hsu}, C.-C., {Burgasser}, A.~J., {Theissen}, C.~A., {et~al.} 2021, \apjs, 257, 45, \dodoi{10.3847/1538-4365/ac1c7d}

\bibitem[{{Hsu} {et~al.}(2024{\natexlab{a}}){Hsu}, {Wang}, {Xuan}, {Ruffio}, {Morris}, {Echeverri}, {Xin}, {Liberman}, {Finnerty}, {Horstman}, {Sappey}, {Doppmann}, {Mawet}, {Jovanovic}, {Fitzgerald}, {Delorme}, {Wallace}, {Baker}, {Bartos}, {Blake}, {Calvin}, {Cetre}, {L{\'o}pez}, {Pezzato}, {Schofield}, {Skemer}, \& {Wang}}]{Hsu:2024ab}
{Hsu}, C.-C., {Wang}, J.~J., {Xuan}, J.~W., {et~al.} 2024{\natexlab{a}}, \apj, 971, 9, \dodoi{10.3847/1538-4357/ad58d3}

\bibitem[{{Hsu} {et~al.}(2024{\natexlab{b}}){Hsu}, {Burgasser}, {Theissen}, {Birky}, {Aganze}, {Gerasimov}, {Schmidt}, {Blake}, {Covey}, {Moreno-Hilario}, {Gelino}, {Serna}, {Brownstein}, \& {Cunha}}]{Hsu:2024aa}
{Hsu}, C.-C., {Burgasser}, A.~J., {Theissen}, C.~A., {et~al.} 2024{\natexlab{b}}, \apjs, 274, 40, \dodoi{10.3847/1538-4365/ad6b27}

\bibitem[{{Hunter}(2007)}]{Hunter:2007aa}
{Hunter}, J.~D. 2007, Computing in Science and Engineering, 9, 90, \dodoi{10.1109/MCSE.2007.55}

\bibitem[{Jeffreys(1961)}]{Jeffreys:1961aa}
Jeffreys, H. 1961, Theory of Probability, 3rd ed. (Oxford Classic Texts in the Physical Sciences. Oxford Univ. Press)

\bibitem[{{Jiang} {et~al.}(2023){Jiang}, {Wang}, {Ormel}, {Krijt}, \& {Dong}}]{Jiang:2023aa}
{Jiang}, H., {Wang}, Y., {Ormel}, C.~W., {Krijt}, S., \& {Dong}, R. 2023, \aap, 678, A33, \dodoi{10.1051/0004-6361/202346637}

\bibitem[{{Keenan}(1993)}]{Keenan:1993aa}
{Keenan}, P.~C. 1993, \pasp, 105, 905, \dodoi{10.1086/133252}

\bibitem[{{Keppler} {et~al.}(2018){Keppler}, {Benisty}, {M{\"u}ller}, {Henning}, {van Boekel}, {Cantalloube}, {Ginski}, {van Holstein}, {Maire}, {Pohl}, {Samland}, {Avenhaus}, {Baudino}, {Boccaletti}, {de Boer}, {Bonnefoy}, {Chauvin}, {Desidera}, {Langlois}, {Lazzoni}, {Marleau}, {Mordasini}, {Pawellek}, {Stolker}, {Vigan}, {Zurlo}, {Birnstiel}, {Brandner}, {Feldt}, {Flock}, {Girard}, {Gratton}, {Hagelberg}, {Isella}, {Janson}, {Juhasz}, {Kemmer}, {Kral}, {Lagrange}, {Launhardt}, {Matter}, {M{\'e}nard}, {Milli}, {Molli{\`e}re}, {Olofsson}, {P{\'e}rez}, {Pinilla}, {Pinte}, {Quanz}, {Schmidt}, {Udry}, {Wahhaj}, {Williams}, {Buenzli}, {Cudel}, {Dominik}, {Galicher}, {Kasper}, {Lannier}, {Mesa}, {Mouillet}, {Peretti}, {Perrot}, {Salter}, {Sissa}, {Wildi}, {Abe}, {Antichi}, {Augereau}, {Baruffolo}, {Baudoz}, {Bazzon}, {Beuzit}, {Blanchard}, {Brems}, {Buey}, {De Caprio}, {Carbillet}, {Carle}, {Cascone}, {Cheetham}, {Claudi}, {Costille}, {Delboulb{\'e}}, {Dohlen}, {Fantinel}, {Feautrier}, {Fusco}, {Giro}, {Gluck},
  {Gry}, {Hubin}, {Hugot}, {Jaquet}, {Le Mignant}, {Llored}, {Madec}, {Magnard}, {Martinez}, {Maurel}, {Meyer}, {M{\"o}ller-Nilsson}, {Moulin}, {Mugnier}, {Orign{\'e}}, {Pavlov}, {Perret}, {Petit}, {Pragt}, {Puget}, {Rabou}, {Ramos}, {Rigal}, {Rochat}, {Roelfsema}, {Rousset}, {Roux}, {Salasnich}, {Sauvage}, {Sevin}, {Soenke}, {Stadler}, {Suarez}, {Turatto}, \& {Weber}}]{Keppler:2018aa}
{Keppler}, M., {Benisty}, M., {M{\"u}ller}, A., {et~al.} 2018, \aap, 617, A44, \dodoi{10.1051/0004-6361/201832957}

\bibitem[{Konopacky {et~al.}(2023)Konopacky, Baker, Mawet, Fitzgerald, Jovanovic, Beichman, Ruane, Bertz, Terada, Dekany, Lingvay, Kassis, Anderson, Tamura, Benneke, Beatty, Do, Nishiyama, Plavchan, Wang, Wang, Burgasser, Ruffio, Zhang, Brown, Fucik, Gibbs, Gibson, Halverson, Johnson, Karkar, Kotani, Kress, Leifer, Magnone, Maire, Pahuja, Porter, Roberts, Sappey, Thorne, Wang, Artigau, Blake, Canalizo, Chen, Doppmann, Doyon, Dressing, Fang, Greene, Herczeg, Hillenbrand, Howard, Kane, Kataria, Kempton, Knutson, Lafreni{\`e}re, Liu, Metchev, Millar-Blanchaer, Narita, Pandey, Rajaguru, Robertson, Salyk, Sato, Schlawin, Sengupta, Sivarani, Skidmore, Vasisht, Yasui, \& Zhang}]{Konopacky:2023aa}
Konopacky, Q.~M., Baker, A.~D., Mawet, D., {et~al.} 2023, in Techniques and Instrumentation for Detection of Exoplanets XI, ed. G.~J. Ruane, Vol. 12680, International Society for Optics and Photonics (SPIE), 1268007, \dodoi{10.1117/12.2681522}

\bibitem[{{Kotani} {et~al.}(2020){Kotani}, {Kawahara}, {Ishizuka}, {Jovanovic}, {Vievard}, {Lozi}, {Sahoo}, {Guyon}, {Yoneta}, \& {Tamura}}]{Kotani:2020aa}
{Kotani}, T., {Kawahara}, H., {Ishizuka}, M., {et~al.} 2020, in Society of Photo-Optical Instrumentation Engineers (SPIE) Conference Series, Vol. 11448, Adaptive Optics Systems VII, ed. L.~{Schreiber}, D.~{Schmidt}, \& E.~{Vernet}, 1144878, \dodoi{10.1117/12.2561755}

\bibitem[{{Law} {et~al.}(2024){Law}, {Benisty}, {Facchini}, {Teague}, {Bae}, {Isella}, {Kamp}, {{\"O}berg}, {Portilla-Revelo}, \& {Rampinelli}}]{Law:2024aa}
{Law}, C.~J., {Benisty}, M., {Facchini}, S., {et~al.} 2024, \apj, 964, 190, \dodoi{10.3847/1538-4357/ad24d2}

\bibitem[{Marley {et~al.}(2018)Marley, Saumon, Morley, \& Fortney}]{Marley:2018aa}
Marley, M., Saumon, D., Morley, C., \& Fortney, J. 2018, {Sonora 2018: Cloud-free, solar composition, solar C/O substellar evolution models}, 1.0,  Zenodo, \dodoi{10.5281/zenodo.2628068}

\bibitem[{{Marley} {et~al.}(2021){Marley}, {Saumon}, {Visscher}, {Lupu}, {Freedman}, {Morley}, {Fortney}, {Seay}, {Smith}, {Teal}, \& {Wang}}]{Marley:2021aa}
{Marley}, M.~S., {Saumon}, D., {Visscher}, C., {et~al.} 2021, \apj, 920, 85, \dodoi{10.3847/1538-4357/ac141d}

\bibitem[{{Martin} {et~al.}(2018){Martin}, {Fitzgerald}, {McLean}, {Doppmann}, {Kassis}, {Aliado}, {Canfield}, {Johnson}, {Kress}, {Lanclos}, {Magnone}, {Sohn}, {Wang}, \& {Weiss}}]{Martin:2018aa}
{Martin}, E.~C., {Fitzgerald}, M.~P., {McLean}, I.~S., {et~al.} 2018, in Society of Photo-Optical Instrumentation Engineers (SPIE) Conference Series, Vol. 10702, Ground-based and Airborne Instrumentation for Astronomy VII, ed. C.~J. {Evans}, L.~{Simard}, \& H.~{Takami}, 107020A, \dodoi{10.1117/12.2312266}

\bibitem[{{Mawet} {et~al.}(2016){Mawet}, {Wizinowich}, {Dekany}, {Chun}, {Hall}, {Cetre}, {Guyon}, {Wallace}, {Bowler}, {Liu}, {Ruane}, {Serabyn}, {Bartos}, {Wang}, {Vasisht}, {Fitzgerald}, {Skemer}, {Ireland}, {Fucik}, {Fortney}, {Crossfield}, {Hu}, \& {Benneke}}]{Mawet:2016aa}
{Mawet}, D., {Wizinowich}, P., {Dekany}, R., {et~al.} 2016, in Society of Photo-Optical Instrumentation Engineers (SPIE) Conference Series, Vol. 9909, Adaptive Optics Systems V, ed. E.~{Marchetti}, L.~M. {Close}, \& J.-P. {V{\'e}ran}, 99090D, \dodoi{10.1117/12.2233658}

\bibitem[{{Mawet} {et~al.}(2017){Mawet}, {Delorme}, {Jovanovic}, {Wallace}, {Bartos}, {Wizinowich}, {Fitzgerald}, {Lilley}, {Ruane}, {Wang}, {Klimovich}, \& {Xin}}]{Mawet:2017aa}
{Mawet}, D., {Delorme}, J.~R., {Jovanovic}, N., {et~al.} 2017, in Society of Photo-Optical Instrumentation Engineers (SPIE) Conference Series, Vol. 10400, Society of Photo-Optical Instrumentation Engineers (SPIE) Conference Series, ed. S.~{Shaklan}, 1040029, \dodoi{10.1117/12.2274891}

\bibitem[{{Mawet} {et~al.}(2019){Mawet}, {Fitzgerald}, {Konopacky}, {Beichman}, {Jovanovic}, {Dekany}, {Hover}, {Chisholm}, {Ciardi}, {Artigau}, {Banyal}, {Beatty}, {Benneke}, {Blake}, {Burgasser}, {Canalizo}, {Chen}, {Do}, {Doppmann}, {Doyon}, {Dressing}, {Fang}, {Greene}, {Hillenbrand}, {Howard}, {Kane}, {Kataria}, {Kempton}, {Knutson}, {Kotani}, {Lafreni{\`e}re}, {Liu}, {Nishiyama}, {Pandey}, {Plavchan}, {Prato}, {Rajaguru}, {Robertson}, {Salyk}, {Sato}, {Schlawin}, {Sengupta}, {Sivarani}, {Skidmore}, {Tamura}, {Terada}, {Vasisht}, {Wang}, \& {Zhang}}]{Mawet:2019aa}
{Mawet}, D., {Fitzgerald}, M., {Konopacky}, Q., {et~al.} 2019, in Bulletin of the American Astronomical Society, Vol.~51, 134

\bibitem[{{Mayor} \& {Queloz}(1995)}]{Mayor:1995aa}
{Mayor}, M., \& {Queloz}, D. 1995, \nat, 378, 355, \dodoi{10.1038/378355a0}

\bibitem[{{McLean} {et~al.}(2000){McLean}, {Graham}, {Becklin}, {Figer}, {Larkin}, {Levenson}, \& {Teplitz}}]{McLean:2000aa}
{McLean}, I.~S., {Graham}, J.~R., {Becklin}, E.~E., {et~al.} 2000, in Society of Photo-Optical Instrumentation Engineers (SPIE) Conference Series, Vol. 4008, Optical and IR Telescope Instrumentation and Detectors, ed. M.~{Iye} \& A.~F. {Moorwood}, 1048--1055, \dodoi{10.1117/12.395422}

\bibitem[{{McLean} {et~al.}(1998){McLean}, {Becklin}, {Bendiksen}, {Brims}, {Canfield}, {Figer}, {Graham}, {Hare}, {Lacayanga}, {Larkin}, {Larson}, {Levenson}, {Magnone}, {Teplitz}, \& {Wong}}]{McLean:1998aa}
{McLean}, I.~S., {Becklin}, E.~E., {Bendiksen}, O., {et~al.} 1998, in Society of Photo-Optical Instrumentation Engineers (SPIE) Conference Series, Vol. 3354, Infrared Astronomical Instrumentation, ed. A.~M. {Fowler}, 566--578, \dodoi{10.1117/12.317283}

\bibitem[{{Moehler} {et~al.}(2014){Moehler}, {Modigliani}, {Freudling}, {Giammichele}, {Gianninas}, {Gonneau}, {Kausch}, {Lan{\c{c}}on}, {Noll}, {Rauch}, \& {Vinther}}]{Moehler:2014aa}
{Moehler}, S., {Modigliani}, A., {Freudling}, W., {et~al.} 2014, \aap, 568, A9, \dodoi{10.1051/0004-6361/201423790}

\bibitem[{{Molli{\`e}re} {et~al.}(2019){Molli{\`e}re}, {Wardenier}, {van Boekel}, {Henning}, {Molaverdikhani}, \& {Snellen}}]{Molliere:2019ab}
{Molli{\`e}re}, P., {Wardenier}, J.~P., {van Boekel}, R., {et~al.} 2019, \aap, 627, A67, \dodoi{10.1051/0004-6361/201935470}

\bibitem[{{Molli{\`e}re} {et~al.}(2020){Molli{\`e}re}, {Stolker}, {Lacour}, {Otten}, {Shangguan}, {Charnay}, {Molyarova}, {Nowak}, {Henning}, {Marleau}, {Semenov}, {van Dishoeck}, {Eisenhauer}, {Garcia}, {Garcia Lopez}, {Girard}, {Greenbaum}, {Hinkley}, {Kervella}, {Kreidberg}, {Maire}, {Nasedkin}, {Pueyo}, {Snellen}, {Vigan}, {Wang}, {de Zeeuw}, \& {Zurlo}}]{Molliere:2020aa}
{Molli{\`e}re}, P., {Stolker}, T., {Lacour}, S., {et~al.} 2020, \aap, 640, A131, \dodoi{10.1051/0004-6361/202038325}

\bibitem[{{Mu{\~n}oz-Romero} {et~al.}(2024){Mu{\~n}oz-Romero}, {Banzatti}, {{\"O}berg}, {Pontoppidan}, {Salyk}, {Najita}, {Blake}, {Krijt}, {Arulanantham}, {Pinilla}, {Long}, {Rosotti}, {Andrews}, {Wilner}, {Calahan}, \& {The JDISCS Collaboration}}]{Munoz-Romero24}
{Mu{\~n}oz-Romero}, C.~E., {Banzatti}, A., {{\"O}berg}, K.~I., {et~al.} 2024, arXiv e-prints, arXiv:2409.03831, \dodoi{10.48550/arXiv.2409.03831}

\bibitem[{{M{\"u}ller} {et~al.}(2018){M{\"u}ller}, {Keppler}, {Henning}, {Samland}, {Chauvin}, {Beust}, {Maire}, {Molaverdikhani}, {van Boekel}, {Benisty}, {Boccaletti}, {Bonnefoy}, {Cantalloube}, {Charnay}, {Baudino}, {Gennaro}, {Long}, {Cheetham}, {Desidera}, {Feldt}, {Fusco}, {Girard}, {Gratton}, {Hagelberg}, {Janson}, {Lagrange}, {Langlois}, {Lazzoni}, {Ligi}, {M{\'e}nard}, {Mesa}, {Meyer}, {Molli{\`e}re}, {Mordasini}, {Moulin}, {Pavlov}, {Pawellek}, {Quanz}, {Ramos}, {Rouan}, {Sissa}, {Stadler}, {Vigan}, {Wahhaj}, {Weber}, \& {Zurlo}}]{Muller:2018aa}
{M{\"u}ller}, A., {Keppler}, M., {Henning}, T., {et~al.} 2018, \aap, 617, L2, \dodoi{10.1051/0004-6361/201833584}

\bibitem[{{{\"O}berg} {et~al.}(2011){{\"O}berg}, {Murray-Clay}, \& {Bergin}}]{Oberg:2011ab}
{{\"O}berg}, K.~I., {Murray-Clay}, R., \& {Bergin}, E.~A. 2011, \apjl, 743, L16, \dodoi{10.1088/2041-8205/743/1/L16}

\bibitem[{{Otten} {et~al.}(2021){Otten}, {Vigan}, {Muslimov}, {N'Diaye}, {Choquet}, {Seemann}, {Dohlen}, {Houll{\'e}}, {Cristofari}, {Phillips}, {Charles}, {Baraffe}, {Beuzit}, {Costille}, {Dorn}, {El Morsy}, {Kasper}, {Lopez}, {Mordasini}, {Pourcelot}, {Reiners}, \& {Sauvage}}]{Otten:2021aa}
{Otten}, G.~P.~P.~L., {Vigan}, A., {Muslimov}, E., {et~al.} 2021, \aap, 646, A150, \dodoi{10.1051/0004-6361/202038517}

\bibitem[{{Pascucci} {et~al.}(2013){Pascucci}, {Herczeg}, {Carr}, \& {Bruderer}}]{Pascucci13}
{Pascucci}, I., {Herczeg}, G., {Carr}, J.~S., \& {Bruderer}, S. 2013, \apj, 779, 178, \dodoi{10.1088/0004-637X/779/2/178}

\bibitem[{{Pascucci} {et~al.}(2023){Pascucci}, {Skinner}, {Deng}, {Ruaud}, {Gorti}, {Schwarz}, {Chapillon}, {Vioque}, \& {Miley}}]{Pascucci:2023aa}
{Pascucci}, I., {Skinner}, B.~N., {Deng}, D., {et~al.} 2023, \apj, 953, 183, \dodoi{10.3847/1538-4357/ace4bf}

\bibitem[{{Pecaut} \& {Mamajek}(2016)}]{Pecaut:2016aa}
{Pecaut}, M.~J., \& {Mamajek}, E.~E. 2016, \mnras, 461, 794, \dodoi{10.1093/mnras/stw1300}

\bibitem[{{Perotti} {et~al.}(2023){Perotti}, {Christiaens}, {Henning}, {Tabone}, {Waters}, {Kamp}, {Olofsson}, {Grant}, {Gasman}, {Bouwman}, {Samland}, {Franceschi}, {van Dishoeck}, {Schwarz}, {G{\"u}del}, {Lagage}, {Ray}, {Vandenbussche}, {Abergel}, {Absil}, {Arabhavi}, {Argyriou}, {Barrado}, {Boccaletti}, {Caratti o Garatti}, {Geers}, {Glauser}, {Justannont}, {Lahuis}, {Mueller}, {Nehm{\'e}}, {Pantin}, {Scheithauer}, {Waelkens}, {Guadarrama}, {Jang}, {Kanwar}, {Morales-Calder{\'o}n}, {Pawellek}, {Rodgers-Lee}, {Schreiber}, {Colina}, {Greve}, {{\"O}stlin}, \& {Wright}}]{Perotti:2023aa}
{Perotti}, G., {Christiaens}, V., {Henning}, T., {et~al.} 2023, \nat, 620, 516, \dodoi{10.1038/s41586-023-06317-9}

\bibitem[{{Pfalzner} \& {Dincer}(2024)}]{Pfalzner:2024aa}
{Pfalzner}, S., \& {Dincer}, F. 2024, \apj, 963, 122, \dodoi{10.3847/1538-4357/ad1bef}

\bibitem[{{Pontoppidan} {et~al.}(2024){Pontoppidan}, {Salyk}, {Banzatti}, {Zhang}, {Pascucci}, {{\"O}berg}, {Long}, {Romero-Mirza}, {Carr}, {Najita}, {Blake}, {Arulanantham}, {Andrews}, {Ballering}, {Bergin}, {Calahan}, {Cobb}, {Colmenares}, {Dickson-Vandervelde}, {Dignan}, {Green}, {Heretz}, {Herczeg}, {Kalyaan}, {Krijt}, {Pauly}, {Pinilla}, {Trapman}, \& {Xie}}]{Pontoppidan24}
{Pontoppidan}, K.~M., {Salyk}, C., {Banzatti}, A., {et~al.} 2024, \apj, 963, 158, \dodoi{10.3847/1538-4357/ad20f0}

\bibitem[{{Ratzenb{\"o}ck} {et~al.}(2023{\natexlab{a}}){Ratzenb{\"o}ck}, {Gro{\ss}schedl}, {M{\"o}ller}, {Alves}, {Bomze}, \& {Meingast}}]{Ratzenbock:2023ab}
{Ratzenb{\"o}ck}, S., {Gro{\ss}schedl}, J.~E., {M{\"o}ller}, T., {et~al.} 2023{\natexlab{a}}, \aap, 677, A59, \dodoi{10.1051/0004-6361/202243690}

\bibitem[{{Ratzenb{\"o}ck} {et~al.}(2023{\natexlab{b}}){Ratzenb{\"o}ck}, {Gro{\ss}schedl}, {Alves}, {Miret-Roig}, {Bomze}, {Forbes}, {Goodman}, {Hacar}, {Lin}, {Meingast}, {M{\"o}ller}, {Piecka}, {Posch}, {Rottensteiner}, {Swiggum}, \& {Zucker}}]{Ratzenbock:2023aa}
{Ratzenb{\"o}ck}, S., {Gro{\ss}schedl}, J.~E., {Alves}, J., {et~al.} 2023{\natexlab{b}}, \aap, 678, A71, \dodoi{10.1051/0004-6361/202346901}

\bibitem[{{Riaud} {et~al.}(2006){Riaud}, {Mawet}, {Absil}, {Boccaletti}, {Baudoz}, {Herwats}, \& {Surdej}}]{Riaud:2006aa}
{Riaud}, P., {Mawet}, D., {Absil}, O., {et~al.} 2006, \aap, 458, 317, \dodoi{10.1051/0004-6361:20065232}

\bibitem[{{Rothman} {et~al.}(2010){Rothman}, {Gordon}, {Barber}, {Dothe}, {Gamache}, {Goldman}, {Perevalov}, {Tashkun}, \& {Tennyson}}]{Rothman:2010aa}
{Rothman}, L.~S., {Gordon}, I.~E., {Barber}, R.~J., {et~al.} 2010, \jqsrt, 111, 2139, \dodoi{10.1016/j.jqsrt.2010.05.001}

\bibitem[{{Ruffio} {et~al.}(2023){Ruffio}, {Horstman}, {Mawet}, {Rosenthal}, {Batygin}, {Wang}, {Millar-Blanchaer}, {Wang}, {Fulton}, {Konopacky}, {Agrawal}, {Hirsch}, {Howard}, {Blunt}, {Nielsen}, {Baker}, {Bartos}, {Bond}, {Calvin}, {Cetre}, {Delorme}, {Doppmann}, {Echeverri}, {Finnerty}, {Fitzgerald}, {Jovanovic}, {L{\'o}pez}, {Martin}, {Morris}, {Pezzato}, {Ruane}, {Sappey}, {Schofield}, {Skemer}, {Venenciano}, {Wallace}, {Wallack}, {Wizinowich}, \& {Xuan}}]{Ruffio:2023aa}
{Ruffio}, J.-B., {Horstman}, K., {Mawet}, D., {et~al.} 2023, \aj, 165, 113, \dodoi{10.3847/1538-3881/acb34a}

\bibitem[{{Speagle}(2020)}]{Speagle:2020aa}
{Speagle}, J.~S. 2020, \mnras, 493, 3132, \dodoi{10.1093/mnras/staa278}

\bibitem[{{Steinmetz} {et~al.}(2020{\natexlab{a}}){Steinmetz}, {Matijevi{\v{c}}}, {Enke}, {Zwitter}, {Guiglion}, {McMillan}, {Kordopatis}, {Valentini}, {Chiappini}, {Casagrande}, {Wojno}, {Anguiano}, {Bienaym{\'e}}, {Bijaoui}, {Binney}, {Burton}, {Cass}, {de Laverny}, {Fiegert}, {Freeman}, {Fulbright}, {Gibson}, {Gilmore}, {Grebel}, {Helmi}, {Kunder}, {Munari}, {Navarro}, {Parker}, {Ruchti}, {Recio-Blanco}, {Reid}, {Seabroke}, {Siviero}, {Siebert}, {Stupar}, {Watson}, {Williams}, {Wyse}, {Anders}, {Antoja}, {Birko}, {Bland-Hawthorn}, {Bossini}, {Garc{\'\i}a}, {Carrillo}, {Chaplin}, {Elsworth}, {Famaey}, {Gerhard}, {Jofre}, {Just}, {Mathur}, {Miglio}, {Minchev}, {Monari}, {Mosser}, {Ritter}, {Rodrigues}, {Scholz}, {Sharma}, {Sysoliatina}, \& {RAVE Collaboration}}]{Steinmetz:2020aa}
{Steinmetz}, M., {Matijevi{\v{c}}}, G., {Enke}, H., {et~al.} 2020{\natexlab{a}}, \aj, 160, 82, \dodoi{10.3847/1538-3881/ab9ab9}

\bibitem[{{Steinmetz} {et~al.}(2020{\natexlab{b}}){Steinmetz}, {Guiglion}, {McMillan}, {Matijevi{\v{c}}}, {Enke}, {Kordopatis}, {Zwitter}, {Valentini}, {Chiappini}, {Casagrande}, {Wojno}, {Anguiano}, {Bienaym{\'e}}, {Bijaoui}, {Binney}, {Burton}, {Cass}, {de Laverny}, {Fiegert}, {Freeman}, {Fulbright}, {Gibson}, {Gilmore}, {Grebel}, {Helmi}, {Kunder}, {Munari}, {Navarro}, {Parker}, {Ruchti}, {Recio-Blanco}, {Reid}, {Seabroke}, {Siviero}, {Siebert}, {Stupar}, {Watson}, {Williams}, {Wyse}, {Anders}, {Antoja}, {Birko}, {Bland-Hawthorn}, {Bossini}, {Garc{\'\i}a}, {Carrillo}, {Chaplin}, {Elsworth}, {Famaey}, {Gerhard}, {Jofre}, {Just}, {Mathur}, {Miglio}, {Minchev}, {Monari}, {Mosser}, {Ritter}, {Rodrigues}, {Scholz}, {Sharma}, {Sysoliatina}, \& {RAVE Collaboration}}]{Steinmetz:2020ab}
{Steinmetz}, M., {Guiglion}, G., {McMillan}, P.~J., {et~al.} 2020{\natexlab{b}}, \aj, 160, 83, \dodoi{10.3847/1538-3881/ab9ab8}

\bibitem[{{Su{\'a}rez} \& {Metchev}(2022)}]{Suarez:2022aa}
{Su{\'a}rez}, G., \& {Metchev}, S. 2022, \mnras, 513, 5701, \dodoi{10.1093/mnras/stac1205}

\bibitem[{{Swastik} {et~al.}(2021){Swastik}, {Banyal}, {Narang}, {Manoj}, {Sivarani}, {Reddy}, \& {Rajaguru}}]{Swastik:2021aa}
{Swastik}, C., {Banyal}, R.~K., {Narang}, M., {et~al.} 2021, \aj, 161, 114, \dodoi{10.3847/1538-3881/abd802}

\bibitem[{{Tabone} {et~al.}(2023){Tabone}, {Bettoni}, {van Dishoeck}, {Arabhavi}, {Grant}, {Gasman}, {Henning}, {Kamp}, {G{\"u}del}, {Lagage}, {Ray}, {Vandenbussche}, {Abergel}, {Absil}, {Argyriou}, {Barrado}, {Boccaletti}, {Bouwman}, {Caratti o Garatti}, {Geers}, {Glauser}, {Justannont}, {Lahuis}, {Mueller}, {Nehm{\'e}}, {Olofsson}, {Pantin}, {Scheithauer}, {Waelkens}, {Waters}, {Black}, {Christiaens}, {Guadarrama}, {Morales-Calder{\'o}n}, {Jang}, {Kanwar}, {Pawellek}, {Perotti}, {Perrin}, {Rodgers-Lee}, {Samland}, {Schreiber}, {Schwarz}, {Colina}, {{\"O}stlin}, \& {Wright}}]{Tabone23}
{Tabone}, B., {Bettoni}, G., {van Dishoeck}, E.~F., {et~al.} 2023, Nature Astronomy, 7, 805, \dodoi{10.1038/s41550-023-01965-3}

\bibitem[{{Thanathibodee} {et~al.}(2020){Thanathibodee}, {Molina}, {Calvet}, {Serna}, {Bae}, {Reynolds}, {Hern{\'a}ndez}, {Muzerolle}, \& {Hern{\'a}ndez}}]{Thanathibodee:2020aa}
{Thanathibodee}, T., {Molina}, B., {Calvet}, N., {et~al.} 2020, \apj, 892, 81, \dodoi{10.3847/1538-4357/ab77c1}

\bibitem[{{Theissen} {et~al.}(2022){Theissen}, {Konopacky}, {Lu}, {Kim}, {Zhang}, {Hsu}, {Chu}, \& {Wei}}]{Theissen:2022aa}
{Theissen}, C.~A., {Konopacky}, Q.~M., {Lu}, J.~R., {et~al.} 2022, \apj, 926, 141, \dodoi{10.3847/1538-4357/ac3252}

\bibitem[{Virtanen {et~al.}(2020)Virtanen, Gommers, Oliphant, Haberland, Reddy, Cournapeau, Burovski, Peterson, Weckesser, Bright, {van der Walt}, Brett, Wilson, Millman, Mayorov, Nelson, Jones, Kern, Larson, Carey, Polat, Feng, Moore, {VanderPlas}, Laxalde, Perktold, Cimrman, Henriksen, Quintero, Harris, Archibald, Ribeiro, Pedregosa, {van Mulbregt}, \& {SciPy 1.0 Contributors}}]{Virtanen:2020aa}
Virtanen, P., Gommers, R., Oliphant, T.~E., {et~al.} 2020, Nature Methods, 17, 261, \dodoi{10.1038/s41592-019-0686-2}

\bibitem[{{Wang} {et~al.}(2021{\natexlab{a}}){Wang}, {Kulikauskas}, \& {Blunt}}]{Wang:2021ac}
{Wang}, J.~J., {Kulikauskas}, M., \& {Blunt}, S. 2021{\natexlab{a}}, {whereistheplanet: Predicting positions of directly imaged companions}, Astrophysics Source Code Library, record ascl:2101.003

\bibitem[{{Wang} {et~al.}(2020){Wang}, {Ginzburg}, {Ren}, {Wallack}, {Gao}, {Mawet}, {Bond}, {Cetre}, {Wizinowich}, {De Rosa}, {Ruane}, {Liu}, {Absil}, {Alvarez}, {Baranec}, {Choquet}, {Chun}, {Defr{\`e}re}, {Delorme}, {Duch{\^e}ne}, {Forsberg}, {Ghez}, {Guyon}, {Hall}, {Huby}, {Jolivet}, {Jensen-Clem}, {Jovanovic}, {Karlsson}, {Lilley}, {Matthews}, {M{\'e}nard}, {Meshkat}, {Millar-Blanchaer}, {Ngo}, {Orban de Xivry}, {Pinte}, {Ragland}, {Serabyn}, {Catal{\'a}n}, {Wang}, {Wetherell}, {Williams}, {Ygouf}, \& {Zuckerman}}]{Wang:2020aa}
{Wang}, J.~J., {Ginzburg}, S., {Ren}, B., {et~al.} 2020, \aj, 159, 263, \dodoi{10.3847/1538-3881/ab8aef}

\bibitem[{{Wang} {et~al.}(2021{\natexlab{b}}){Wang}, {Vigan}, {Lacour}, {Nowak}, {Stolker}, {De Rosa}, {Ginzburg}, {Gao}, {Abuter}, {Amorim}, {Asensio-Torres}, {Baub{\"o}ck}, {Benisty}, {Berger}, {Beust}, {Beuzit}, {Blunt}, {Boccaletti}, {Bohn}, {Bonnefoy}, {Bonnet}, {Brandner}, {Cantalloube}, {Caselli}, {Charnay}, {Chauvin}, {Choquet}, {Christiaens}, {Cl{\'e}net}, {Coud{\'e} Du Foresto}, {Cridland}, {de Zeeuw}, {Dembet}, {Dexter}, {Drescher}, {Duvert}, {Eckart}, {Eisenhauer}, {Facchini}, {Gao}, {Garcia}, {Garcia Lopez}, {Gardner}, {Gendron}, {Genzel}, {Gillessen}, {Girard}, {Haubois}, {Hei{\ss}el}, {Henning}, {Hinkley}, {Hippler}, {Horrobin}, {Houll{\'e}}, {Hubert}, {Jim{\'e}nez-Rosales}, {Jocou}, {Kammerer}, {Keppler}, {Kervella}, {Meyer}, {Kreidberg}, {Lagrange}, {Lapeyr{\`e}re}, {Le Bouquin}, {L{\'e}na}, {Lutz}, {Maire}, {M{\'e}nard}, {M{\'e}rand}, {Molli{\`e}re}, {Monnier}, {Mouillet}, {M{\"u}ller}, {Nasedkin}, {Ott}, {Otten}, {Paladini}, {Paumard}, {Perraut}, {Perrin}, {Pfuhl}, {Pueyo}, {Rameau},
  {Rodet}, {Rodr{\'\i}guez-Coira}, {Rousset}, {Scheithauer}, {Shangguan}, {Shimizu}, {Stadler}, {Straub}, {Straubmeier}, {Sturm}, {Tacconi}, {van Dishoeck}, {Vincent}, {von Fellenberg}, {Ward-Duong}, {Widmann}, {Wieprecht}, {Wiezorrek}, {Woillez}, \& {Gravity Collaboration}}]{Wang:2021ab}
{Wang}, J.~J., {Vigan}, A., {Lacour}, S., {et~al.} 2021{\natexlab{b}}, \aj, 161, 148, \dodoi{10.3847/1538-3881/abdb2d}

\bibitem[{{Wang} {et~al.}(2021{\natexlab{c}}){Wang}, {Ruffio}, {Morris}, {Delorme}, {Jovanovic}, {Pezzato}, {Echeverri}, {Finnerty}, {Hood}, {Zanazzi}, {Bryan}, {Bond}, {Cetre}, {Martin}, {Mawet}, {Skemer}, {Baker}, {Xuan}, {Wallace}, {Wang}, {Bartos}, {Blake}, {Boden}, {Buzard}, {Calvin}, {Chun}, {Doppmann}, {Dupuy}, {Duch{\^e}ne}, {Feng}, {Fitzgerald}, {Fortney}, {Freedman}, {Knutson}, {Konopacky}, {Lilley}, {Liu}, {Lopez}, {Lupu}, {Marley}, {Meshkat}, {Miles}, {Millar-Blanchaer}, {Ragland}, {Roy}, {Ruane}, {Sappey}, {Schofield}, {Weiss}, {Wetherell}, {Wizinowich}, \& {Ygouf}}]{Wang:2021aa}
{Wang}, J.~J., {Ruffio}, J.-B., {Morris}, E., {et~al.} 2021{\natexlab{c}}, \aj, 162, 148, \dodoi{10.3847/1538-3881/ac1349}

\bibitem[{Waskom(2021)}]{Waskom2021}
Waskom, M.~L. 2021, Journal of Open Source Software, 6, 3021, \dodoi{10.21105/joss.03021}

\bibitem[{{Wells} {et~al.}(2015){Wells}, {Pel}, {Glasse}, {Wright}, {Aitink-Kroes}, {Azzollini}, {Beard}, {Brandl}, {Gallie}, {Geers}, {Glauser}, {Hastings}, {Henning}, {Jager}, {Justtanont}, {Kruizinga}, {Lahuis}, {Lee}, {Martinez-Delgado}, {Mart{\'\i}nez-Galarza}, {Meijers}, {Morrison}, {M{\"u}ller}, {Nakos}, {O'Sullivan}, {Oudenhuysen}, {Parr-Burman}, {Pauwels}, {Rohloff}, {Schmalzl}, {Sykes}, {Thelen}, {van Dishoeck}, {Vandenbussche}, {Venema}, {Visser}, {Waters}, \& {Wright}}]{Wells:2015aa}
{Wells}, M., {Pel}, J.~W., {Glasse}, A., {et~al.} 2015, \pasp, 127, 646, \dodoi{10.1086/682281}

\bibitem[{{Xuan} {et~al.}(2022){Xuan}, {Wang}, {Ruffio}, {Knutson}, {Mawet}, {Molli{\`e}re}, {Kolecki}, {Vigan}, {Mukherjee}, {Wallack}, {Wang}, {Baker}, {Bartos}, {Blake}, {Bond}, {Bryan}, {Calvin}, {Cetre}, {Chun}, {Delorme}, {Doppmann}, {Echeverri}, {Finnerty}, {Fitzgerald}, {Horstman}, {Inglis}, {Jovanovic}, {L{\'o}pez}, {Martin}, {Morris}, {Pezzato}, {Ragland}, {Ren}, {Ruane}, {Sappey}, {Schofield}, {Skemer}, {Venenciano}, {Wallace}, \& {Wizinowich}}]{Xuan:2022aa}
{Xuan}, J.~W., {Wang}, J., {Ruffio}, J.-B., {et~al.} 2022, \apj, 937, 54, \dodoi{10.3847/1538-4357/ac8673}

\bibitem[{{Xuan} {et~al.}(2024{\natexlab{a}}){Xuan}, {Wang}, {Finnerty}, {Horstman}, {Grimm}, {Peck}, {Nielsen}, {Knutson}, {Mawet}, {Isaacson}, {Howard}, {Liu}, {Walker}, {Phillips}, {Blake}, {Ruffio}, {Zhang}, {Inglis}, {Wallack}, {Sanghi}, {Gonzales}, {Dai}, {Baker}, {Bartos}, {Bond}, {Bryan}, {Calvin}, {Cetre}, {Delorme}, {Doppmann}, {Echeverri}, {Fitzgerald}, {Jovanovic}, {Liberman}, {L{\'o}pez}, {Martin}, {Morris}, {Pezzato}, {Ruane}, {Sappey}, {Schofield}, {Skemer}, {Venenciano}, {Wallace}, {Wang}, {Wizinowich}, {Xin}, {Agrawal}, {Do {\'O}}, {Hsu}, \& {Phillips}}]{Xuan:2024aa}
{Xuan}, J.~W., {Wang}, J., {Finnerty}, L., {et~al.} 2024{\natexlab{a}}, \apj, 962, 10, \dodoi{10.3847/1538-4357/ad1243}

\bibitem[{{Xuan} {et~al.}(2024{\natexlab{b}}){Xuan}, {Hsu}, {Finnerty}, {Wang}, {Ruffio}, {Zhang}, {Knutson}, {Mawet}, {Mamajek}, {Inglis}, {Wallack}, {Bryan}, {Blake}, {Molli{\`e}re}, {Hejazi}, {Baker}, {Bartos}, {Calvin}, {Cetre}, {Delorme}, {Doppmann}, {Echeverri}, {Fitzgerald}, {Jovanovic}, {Liberman}, {L{\'o}pez}, {Morris}, {Pezzato}, {Sappey}, {Schofield}, {Skemer}, {Wallace}, {Wang}, {Agrawal}, \& {Horstman}}]{Xuan:2024ab}
{Xuan}, J.~W., {Hsu}, C.-C., {Finnerty}, L., {et~al.} 2024{\natexlab{b}}, \apj, 970, 71, \dodoi{10.3847/1538-4357/ad4796}

\bibitem[{{Yoshida} {et~al.}(2022){Yoshida}, {Nomura}, {Tsukagoshi}, {Furuya}, \& {Ueda}}]{Yoshida:2022aa}
{Yoshida}, T.~C., {Nomura}, H., {Tsukagoshi}, T., {Furuya}, K., \& {Ueda}, T. 2022, \apjl, 937, L14, \dodoi{10.3847/2041-8213/ac903a}

\bibitem[{{Zhang} {et~al.}(2019){Zhang}, {Bergin}, {Schwarz}, {Krijt}, \& {Ciesla}}]{Zhang:2019ac}
{Zhang}, K., {Bergin}, E.~A., {Schwarz}, K., {Krijt}, S., \& {Ciesla}, F. 2019, \apj, 883, 98, \dodoi{10.3847/1538-4357/ab38b9}

\bibitem[{{Zhang} {et~al.}(2021){Zhang}, {Snellen}, {Bohn}, {Molli{\`e}re}, {Ginski}, {Hoeijmakers}, {Kenworthy}, {Mamajek}, {Meshkat}, {Reggiani}, \& {Snik}}]{Zhang:2021ae}
{Zhang}, Y., {Snellen}, I. A.~G., {Bohn}, A.~J., {et~al.} 2021, \nat, 595, 370, \dodoi{10.1038/s41586-021-03616-x}

\end{thebibliography}
\bibliographystyle{aasjournal}

\end{document}